\documentclass[a4paper,11pt]{article}
\pdfoutput=1 

\usepackage{jheppub} 

\usepackage[T1]{fontenc} 

\usepackage{soul}
\usepackage{graphicx} 
\usepackage{dcolumn}
\usepackage{bm}
\usepackage{xcolor}

\usepackage[footnotesize]{caption}
\usepackage{amsmath}
\usepackage{amssymb}
\usepackage[utf8]{inputenc}
\usepackage{mathrsfs}
\usepackage{hyperref}
\usepackage{multirow}
\usepackage{cancel}
\usepackage{enumitem}
\usepackage{float}

\newcommand{\met}{\cancel{E}_T}
\newcommand{\iab}{\rm ab^{-1}}

\title{
Lepton Flavor Violating Dilepton Dijet Signatures from Sterile Neutrinos at Proton Colliders
}

\author[a]{Stefan Antusch}
\author[a]{, Eros Cazzato}
\author[b]{, Oliver Fischer}
\author[a]{, A. Hammad}
\author[c,d]{, Kechen Wang}

\affiliation[a]{Department of Physics, University of Basel, Klingelbergstr. 82, CH-4056 Basel, Switzerland}
\affiliation[b]{Institute for Nuclear Physics, Karlsruhe Institute of Technology, Hermann-von-Helmholtz-Platz 1, D-76344 Eggenstein-Leopoldshafen, Germany}
\affiliation[c]{DESY, Notkestrae 85, D-22607 Hamburg, Germany}
\affiliation[d]{Center for Future High Energy Physics, Institute of High Energy Physics, Chinese Academy of Sciences, Beijing, 100049, China}

\emailAdd{stefan.antusch@unibas.ch}
\emailAdd{e.cazzato@unibas.ch}
\emailAdd{oliver.fischer@kit.edu}
\emailAdd{ahmed.hammad@unibas.ch}
\emailAdd{kechen.wang@desy.de}

\preprint{DESY 17-151}

\abstract{
In this article we investigate the prospects of searching for sterile neutrinos in lowscale seesaw scenarios via the lepton flavour violating (but lepton number conserving) dilepton dijet signature.  In our study, we focus on the final state $e^\pm \mu^\mp jj$ at the HL-LHC and the FCC-hh (or the SppC).
We perform a multivariate analysis
at the detector level
including the dominant SM backgrounds from di-top, di-boson, and tri-boson.
Under the assumption of the active-sterile neutrino mixings $|V_{ l N}|^2=|\theta_e|^2=|\theta_\mu|^2$ and $|V_{ \tau N}|^2 = |\theta_\tau|^2=0$, the sensitivities on the signal production cross section times branching ratio $\sigma(p p \to l^\pm N)\times {\rm BR} (N \to l^{ \mp} jj)$ and on $|V_{ l N}|^2$ for sterile neutrino mass $M_N$ between 200 and 1000 GeV are derived.
For the benchmark $M_N=500$ GeV, when ignoring systematic uncertainties at the HL-LHC (FCC-hh/SppC) with 3 (20) $\iab$ luminosity, the resulting 2-$\sigma$ limits on $|V_{ l N}|^2$ are 
$4.9\times 10^{-3}$
($7.0\times 10^{-5}$), 
while the 2 -$\sigma$ limit on $\sigma \times {\rm BR}$ are 
$4.4\times10^{-2}$ ($1.6\times10^{-2}$)
fb, respectively.
The effect of the systematic uncertainty is also studied and found to be important for sterile neutrinos with 
smaller masses.
We also comment on searches with $\tau^\pm \mu^\mp jj$ and $\tau^\pm e^\mp jj$ final states.
}

\begin{document} 
\maketitle
\flushbottom

\section{Introduction}
\label{sec:intro}
The observation of neutrino oscillations provides evidence that at least two of the involved neutrinos are massive.
The absolute mass scale of the light neutrino masses has not been measured but it is bound to lie below about $0.2$ eV from neutrinoless double beta decay experiments and cosmological constraints, see for instance Ref.~\cite{Gariazzo:2015rra,deGouvea:2015euy} for recent reviews. 
The origin of the neutrino masses is a prominent puzzle of today's elementary particle physics, since it is not possible within the Standard Model (SM) to account for it in a renormalisable way.
Thus neutrino oscillations are evidence from the laboratory for physics beyond the SM.

In the following, we shall focus on the class of SM extensions with neutral fermions, which are gauge singlets and therefore often referred to as ``sterile'' neutrinos, and can provide mass terms for the light neutrinos to explain the observed oscillations.
In particular, the addition of sterile neutrinos allows for a Majorana-type mass term as well as for Dirac-type masses via Yukawa couplings with the SM active neutrino fields.
The sterile and active neutrinos mix when the electroweak symmetry is broken, resulting in light and heavy mass eigenstates. This mass generating mechanism goes by the name of type-I seesaw and is highly searched for by the particle physics community, cf.\ e.g.\ Refs.\ \cite{Aad:2015xaa,Khachatryan:2016olu,Agostini:2017iyd}. Prominent signatures are the likes of neutrinoless double beta decay 
and same-sign dilepton searches at proton colliders.
Furthermore, this class of models can give an explanation for the observed baryon asymmetry of our universe via leptogenesis and of dark matter,
for a recent review see e.g. Ref.\ \cite{Drewes:2013gca} and references therein.

In type-I seesaw, one often assumes either tiny neutrino Yukawa couplings or a very high mass scale for the heavy neutrinos in order to explain the smallness of the light neutrinos' masses. 
This assumption makes it, however, nearly impossible to produce these particles at collider experiments. 

Alternatively, one may impose a protective (``lepton number''-like) symmetry, where a slight breaking from this symmetry is responsible for the small mass of the light neutrinos. 
Various types of symmetry protected seesaw models have been constructed in the literature, cf.\ for instance \cite{Wyler:1982dd,Mohapatra:1986bd,Shaposhnikov:2006nn,Kersten:2007vk,Gavela:2009cd,Malinsky:2005bi}. 
In this framework neither tiny neutrino Yukawa couplings nor large masses for the heavy neutrinos are required to explain the smallness of the light neutrino masses. 
Thus heavy neutrinos with masses around the electroweak scale with unsuppressed Yukawa couplings (and thus unsuppressed active-sterile neutrino mixings) are possible, and their effects can be studied at colliders (cf.\ Ref.\ \cite{Deppisch:2017ecm} for an overview).

Regardless of the underlying model, especially at proton colliders the signatures from sterile neutrinos are often hidden behind comparably enormous rates of SM background for most processes. 
There are a few processes at high-energy colliders where the background does not pose an unsurmountable problem, the most prominent ones being the lepton number violating (LNV) same sign dilepton $\ell^\pm_\alpha\ell^\pm_\alpha$ final states in the dilepton-dijet channel. 
However the signal strength of this type of signature is suppressed together with the LNV by the smallness of the neutrino masses, 
as discussed for instance in Refs.\ \cite{Kersten:2007vk,Antusch:2016ejd,Moffat:2017feq,Antusch:2017ebe}.

On the other hand, as was suggested in Ref.\ \cite{Antusch:2016ejd}, the lepton flavour violating (LFV) (but lepton number conserving (LNC)) dilepton signature, with the final state $\ell_\alpha^\pm\ell^\mp_\beta$ ($\alpha\neq\beta$) (and to some extent also the LFV trilepton signature) has reducible background only while its signal strength is unsuppressed by the light neutrino masses. 

Previous collider studies have focused mostly on same-sign dileptons for the LHC, e.g.~\cite{delAguila:2006bda,Han:2006ip,delAguila:2007qnc,delAguila:2008cj,Atre:2009rg,Chao:2009ef,Das:2012ze,Das:2015toa,Das:2016hof}. 
Some studies of this channel can also be found for future accelerators such as the Future Circular Collider (FCC) \cite{Alva:2014gxa}. 
Also the trilepton channel has gotten attention recently and triggered some studies of LHC discovery prospects \cite{Dib:2016wge,Dib:2017vux,Dib:2017iva}.
Very little attention has been given to the LFV (but LNC) dilepton-dijet channel so far, despite the promising sensitivity obtained from a ``first look'' at the parton level in Refs.\ \cite{Antusch:2016ejd,Arganda:2015ija}.

The goal of this article is therefore to present a thorough investigation of the LFV (but LNC) dilepton-dijet channel as a signature from sterile neutrino extensions of the Standard Model, especially the $e^\pm\mu^\mp jj$ final state.  
Our study goes beyond previous works by discussing relevant backgrounds, performing a fast simulation of the detector response for the signal and background, applying multivariate analysis techniques to separate the signal from the background, as well as including a discussion for the statistical and systematic errors.
We provide sensitivities not only for the high-luminosity Large Hadron Collider (HL-LHC) but also for the FCC in the hadron colliding mode (FCC-hh). Our results are also applicable to the 
Super proton-proton Collider
(SppC)~\cite{Tang:2015qga} depending of course on the final design and the corresponding 
detector performance.

The article is organized as follows. In section~\ref{sec:model}, we briefly describe the theory model we used. In section~\ref{sec:strategy}, we present the search strategy for LFV dilepton-dijet signals from heavy sterile neutrinos. The results at HL-LHC and FCC-hh are shown in section~\ref{sec:results}. We conclude in section~\ref{sec:summary}.

\section{The Theory Model}
\label{sec:model}
We use a specific realisation which captures the relevant features of the symmetry protected seesaw models for the collider phenomenology as our benchmark model. This realisation involves two heavy neutrinos that supposes a ``lepton number''-like symmetry (an extended version of the usual lepton number), which can be found in e.g.~\cite{Antusch:2015mia}. 
For this collider study it is sufficient to focus on the limit of intact protective symmetry, i.e.~symmetry limit, since the signal is lepton number conserving and the light neutrino masses are for collider purposes effectively zero, see below.\footnote{When the symmetry is approximate, viz.~slightly broken, non-degenerate heavy neutrino masses induce LNV.} 

The benchmark model includes one pair of sterile neutrinos $N_R^1$ and $N_R^2$ which are relevant for the collider phenomenology. The resulting Lagrangian density is given by:
\begin{equation}
\mathscr{L} = \mathscr{L}_\mathrm{SM} -  \overline{N_R^1} 
M_N
N^{2\,c}_R - y_{\nu_{\alpha}}\overline{N_{R}^1} \widetilde \phi^\dagger \, L^\alpha
+\mathrm{H.c.} 
+ \dots  \;,
\label{eqn:lagrange}
\end{equation}
where $\mathscr{L}_\mathrm{SM}$ contains the usual SM field content and with $L^\alpha$, $(\alpha=e,\mu,\tau)$, and $\phi$ being the lepton and Higgs doublets, respectively, $y_{\nu_{\alpha}}$ are the complex-valued neutrino Yukawa couplings, and 
$M_N$ 
the sterile neutrino mass. 
The ellipses indicate terms for additional sterile neutrinos which are decoupled from collider phenomenology. 

The symmetric mass matrix ${\cal M}$ of the active and sterile neutrinos is obtained from Eq.~(\ref{eqn:lagrange}) after electroweak symmetry breaking $\mathscr L$ contains $ -1/2\,\overline{n^c} \mathcal{\, M\,} n + H.c$, with $n  =  \left(\nu_{e_L},\nu_{\mu_L},\nu_{\tau_L},(N_R^1)^c,(N_R^2)^c\right)^T$. It can be diagonalized by the unitary 5 $\times$ 5 leptonic mixing matrix $U$:
\begin{equation}
 U^T\, {\cal M}\, U \cong \text{Diag}\left(0,0,0,M_N,M_N\right)  
\,.
\label{eqn:diagonalisation}
\end{equation}
The mass eigenstates $\tilde n_j = \left(\nu_1,\nu_2,\nu_3,N_4,N_5\right)^T_j = U_{j \alpha}^{\dagger} n_\alpha$ are the three light neutrinos, which are massless in the symmetry limit, and two heavy neutrinos with degenerate
mass eigenvalues 
$M_N$
in the symmetric limit. 
The leptonic mixing matrix $U$ in Eq.\ \eqref{eqn:diagonalisation} can be expressed explicitly, cf.\ \cite{Antusch:2015mia}. Its entries are governed by the active-sterile neutrino mixing angles which are quantified via
\begin{equation}
\theta_\alpha = \frac{y_{\nu_\alpha}^{*}}{\sqrt{2}}\frac{v_\mathrm{EW}}{M_N}\,, \qquad |\theta|^2 := \sum_{\alpha} |\theta_\alpha|^2\,,
\label{def:thetaa}
\end{equation}
with  $v_\mathrm{EW} = 246.22$ GeV the vacuum expectation value of the Higgs field. 

Since the light and heavy neutrino mass eigenstates are admixtures of the active and sterile neutrinos, the weak currents, 
cast into the mass basis, are given by
\begin{eqnarray}
j_\mu^\pm & = & \sum\limits_{i=1}^5 \sum\limits_{\alpha=e,\mu,\tau}\frac{g}{\sqrt{2}} \bar \ell_\alpha\, \gamma_\mu\, P_L\, U_{\alpha i}\, \tilde n_i\, + \text{ H.c.}\,, \\
j_\mu^0 & = & \sum\limits_{i,j=1}^5 \sum\limits_{\alpha=e,\mu,\tau}\frac{g}{2\,c_W} \overline{\tilde n_j}\, U^\dagger_{j\alpha}\, \gamma_\mu\, P_L\, U_{\alpha i}\, \tilde n_i\,, 
\label{eqn:weakcurrentmass}
\end{eqnarray}
with $U$ the leptonic mixing matrix, $g$ being the weak coupling constant, $c_W$ the cosine of the Weinberg angle and $P_L = {1 \over 2}(1-\gamma^5)$ the left-chiral projection operator.
The resulting heavy neutrino interactions can be summarized as
\begin{eqnarray}
j_\mu^\pm & \supset &  \frac{g}{2} \, \theta_\alpha \, \bar \ell_\alpha \, \gamma_\mu P_L \left(-\mathrm{i} N_4 + N_5 \right) + \text{H.c.} \,, \label{eqn:weakcurrent1}\\
j_\mu^0 & = & \frac{g}{2\,c_W} \sum\limits_{i,j=1}^5 \vartheta_{ij} \overline{ \tilde n_i} \gamma_\mu P_L \tilde n_j\,, \\
\mathscr{L}_{\rm Yuk.} & \supset & \frac{M_N}{v_\mathrm{EW}} \sum\limits_{i=1}^3 \left(\vartheta_{i4}^* \overline{N_4^c}+ \vartheta_{i5}^*\overline{N^c_5}\right) h\, \nu_i +\text{ H.c.}  \,,
\label{eqn:weakcurrent2}
\end{eqnarray}
with $h = \sqrt{2} \,\mbox{Re}{(\phi^0)}$ being the real scalar Higgs boson and $\vartheta_{ij} =  \sum_{\alpha=e,\mu,\tau} U^\dagger_{i\alpha}U_{\alpha j}^{}$.

In the limit of the protective symmetry being exact, the benchmark model adds seven parameters to the SM, the moduli of the neutrino Yukawa couplings ($|y_{\nu_e}|$, $|y_{\nu_\mu}|$, $|y_{\nu_\tau}|$), their respective phase, or equivalently, the active-sterile mixing angles from Eq.~\eqref{def:thetaa}, and the mass 
$M_N$. The phases may be accessible in neutrino oscillation experiments (see e.g.\ \cite{FernandezMartinez:2007ms,Antusch:2009pm}). 
We restrict ourselves to the four parameters $|\theta_{e}|$, $|\theta_{\mu}|$, $|\theta_{\tau}|$ and 
$M_N$.
In the following, we also use the neutrino mixing matrix elements
$|V_{ \alpha N}|^2$
to present our results, which are commonly used in the literature to quantify the active-sterile neutrino mixing. 
For a fixed flavor $\alpha$ (usually identified via the charged lepton $l_\alpha$) this notation relates to the one introduced above in the following way:
\begin{equation}
|V_{ \alpha N}|^2=|U_{\alpha 4}|^2+|U_{\alpha 5}|^2=|\theta_\alpha |^2\,.
\label{eq:V_lN}
\end{equation}

\section{Search Strategy}
\label{sec:strategy}
Proton colliders provide an environment where the SM can be tested at highest center-of-mass energies.
For what follows we consider the HL-LHC with 14 TeV center-of-mass energy and a total integrated luminosity of 3~ab$^{-1}$\cite{Apollinari:2015bam}. 
We also consider the discussed FCC-hh~\cite{Golling:2016gvc,Mangano:2016jyj,Contino:2016spe} and the 
SppC
~\cite{Tang:2015qga}, with envisaged center-of-mass energies of 100~TeV and target integrated luminosities of around 20~ab$^{-1}$\cite{Hinchliffe:2015qma}. For brevity, we will only refer to the FCC-hh in the following.

\subsection{Signal: Mixed-flavor Dilepton Plus Jets from Heavy Neutrinos}
\label{subsec:signal}

\begin{figure}[h]
\centering
\includegraphics[width=7.5cm,height=5cm]{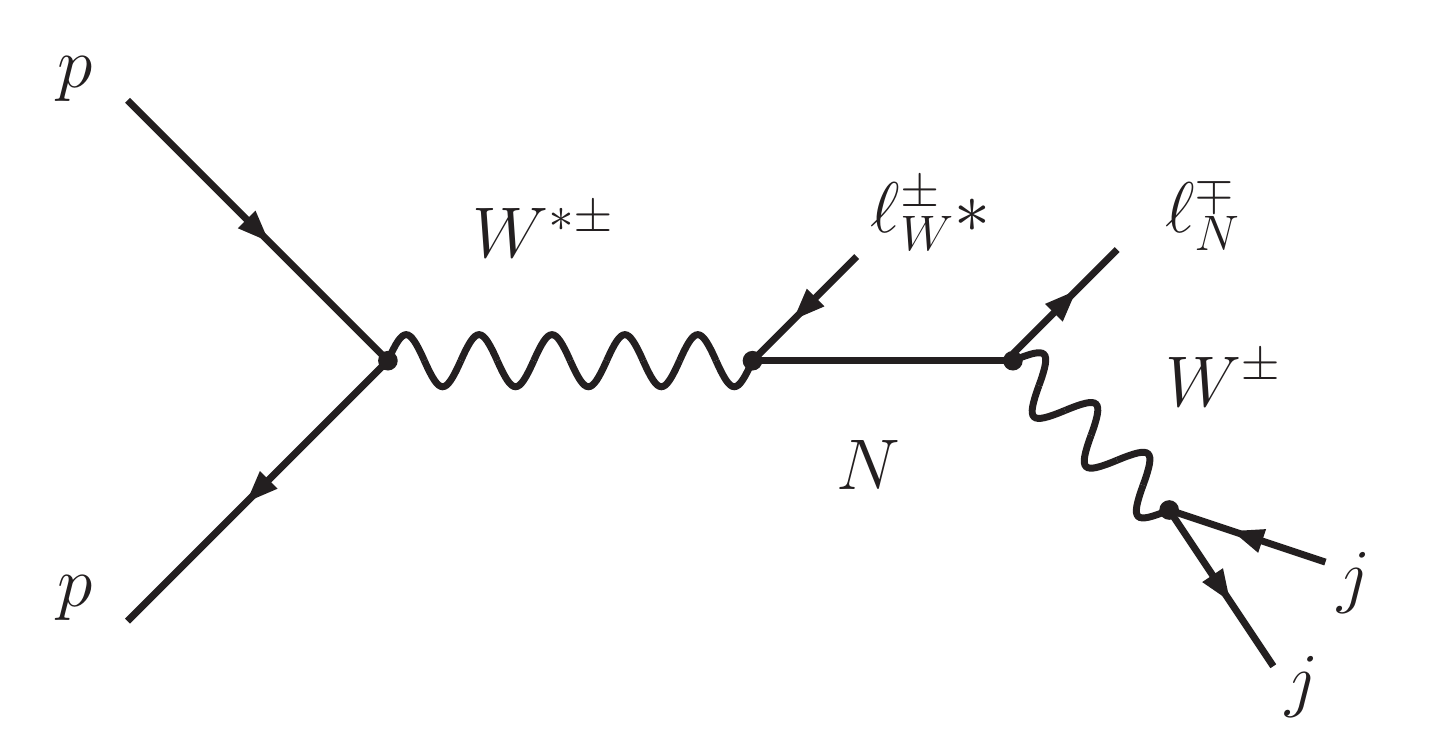}
\caption{
The Feynman diagram depicting the dominant signal production mechanism for heavy neutrino masses and center-of-mass energies as considered in this article.
}
\label{fig:diagram}
\end{figure}

Heavy neutrinos can be produced from proton-proton collisions via Drell-Yan processes, Higgs boson decays, and gauge boson fusion, cf.\ \cite{Degrande:2016aje,Ruiz:2017yyf,Cai:2017mow}.
We focus here on charged current Drell-Yan production of a heavy neutrino with an associated charged lepton yielding $p p \to\ell_\alpha^\pm N$,
cf.\ Fig.\ \ref{fig:diagram}.
It is the dominant production mechanism for heavy neutrino masses around the electroweak scale and the considered center-of-mass energies. 
It is worth noting that the $W\gamma$ fusion production process is the next most important process in the mass range from 200 GeV to $\sim$1 TeV, and it becomes more important and even surpasses the charged current Drell-Yan production for larger heavy neutrino masses \cite{Alva:2014gxa}. 
The contribution from $W\gamma$ adds about 20$\sim$30\% to the LO cross section, cf.\ Ref.~\cite{Alva:2014gxa}.
Due to its limited enhancement on the final discovery limits 
for the here considered mass range,
the $W\gamma$ contributions to the signal are not considered in this study.

As shown in Fig.~\ref{fig:diagram}, the charged current decays of the Drell-Yan produced heavy neutrinos together with the hadronic decay of the final state $W$ boson yield the semileptonic final state $\ell_\alpha^\pm \ell_\beta^\mp j j$.
To discriminate between these two final state leptons, we label the lepton from the Drell-Yan off-shell 
$W^{*\pm}$
as $l_{W^*}^{\pm}$ (i.e. $l_{\alpha}^{\pm}$ or $l^\pm$), while the lepton from the heavy neutrino as $l_N^{\mp}$ (i.e. $l_{\beta}^{\mp}$ or $l^{\prime \mp}$).
We note that for the signal these two leptons 
can
have different flavors.
The event rate is sensitive to the mixing angle combination of $|\theta_\alpha|^2$ and $|\theta_\beta|^2/|\theta|^2$ through the production and decay channel, respectively. 
Here the flavor indices $\alpha,\beta=e,\,\mu,\,\tau$ can be inferred from the charged leptons. 
For $\alpha\neq\beta$, this final state yields a signal for lepton flavour violation, because there is no SM background process at the parton level as discussed in Refs.~\cite{Antusch:2016ejd,Arganda:2015ija}. 
We emphasize that we study the LNC process with leptons of opposite charge since there the signal strength is not suppressed by the smallness of the neutrino masses.\footnote{
Breaking of the protective symmetry can induce LNV by heavy neutrino oscillations 
as discussed in Refs.\ \cite{Gluza:2015goa,Anamiati:2016uxp,Das:2017hmg,Antusch:2017ebe},
but even in a optimistic case the fraction of LNV events is negligible (for $\theta^2\lesssim 10^{-5}$) \cite{Antusch:2017ebe}.
}

The signal for our study is $e^\pm \mu^\mp j j$ with $\alpha=e\, (\mu)$ and $\beta=\mu\, (e)$, which tests
the mixing angle combination $|\theta_e\theta_\mu|^2/|\theta|^2$ or equivalently,
$|V_{ eN}V_{ \mu N}|^2/\sum_{\alpha} |V_{ \alpha N}|^2$.
For practical reasons we make the following assumption and discuss the special case for the active-sterile mixing angles:
\begin{equation}
|V_{ l N}|^2=|\theta_e|^2=|\theta_\mu|^2 \neq 0 \qquad \text{and} \qquad 
|V_{\tau N}|^2 = |\theta_\tau|^2=0\,,
\label{eqn:assumption}
\end{equation}
which implies that $|V_{ eN}V_{ \mu N}|^2/\sum_{\alpha=e, \mu, \tau} |V_{ \alpha N}|^2 = \tfrac{1}{2} |V_{ l N}|^2$.
The results derived below are valid for this case only, but they can be translated to any of the other possible set of active-sterile mixing angles with a numerical overall factor. 

\begin{figure}[htbp]
\centering
\includegraphics[width=7.5cm,height=5cm]{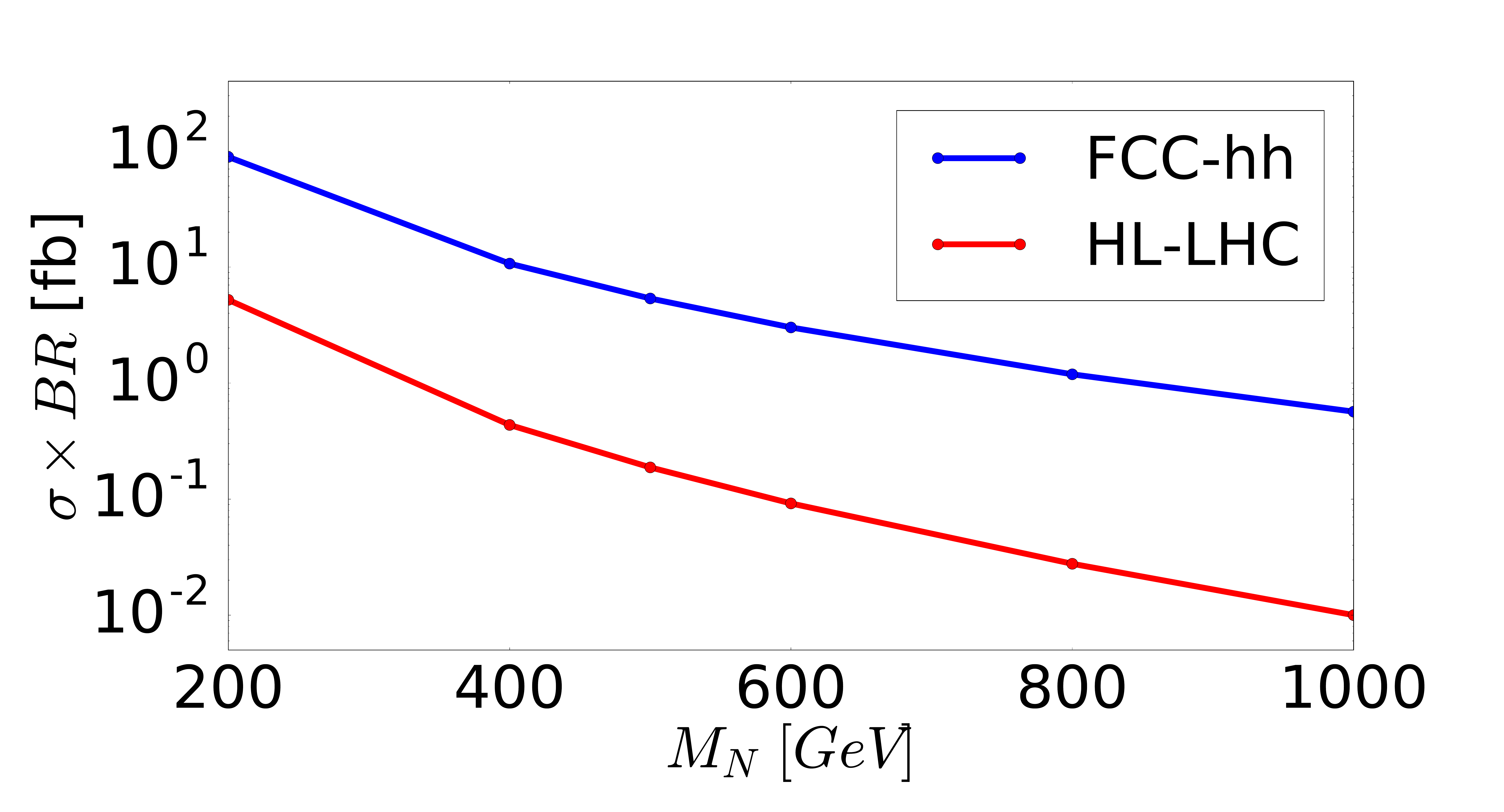}
\caption{Production cross section times branching ratio $\sigma(p p \to l^\pm N)\times {\rm BR} (N \to l^{ \mp} jj)$ in fb for heavy neutrino mass eigenstates via the Drell-Yan processes $pp \to W^* \to \ell^\pm N \to \ell^\pm \ell^\mp jj$ at leading order.
Here, $l = e,\mu$ and the cross section includes di-leptons with all flavor combinations (i.e., $e^\pm \mu^\mp$, $e^+e^-$, and $\mu^+\mu^-$).
The active-sterile mixings are fixed as $|V_{ l N}|^2 = |V_{ e N}|^2 = |V_{ \mu N}|^2 =  10^{-2},\, | V_{\tau N}|=0$.
}
\label{fig:xsection}
\end{figure}

In Fig.~\ref{fig:xsection}, we show the production cross section times branching ratio
$\sigma(p p \to l^\pm N)\times {\rm BR} (N \to l^{\mp} jj)$
in fb at the HL-LHC and FCC-hh when 
$|V_{ lN}|^2 = 10^{-2}$
and $|V_{ \tau N}|=0$. We note that here $l = e,\mu$. Besides the mixed flavor lepton pair $e^\pm \mu^\mp$, the cross sections in this figure also 
include the production of the same flavor lepton pairs $e^+e^-$ and $\mu^+\mu^-$.
The cross sections for a few mass points can be also calculated from the initial number of events in Table~\ref{tab:cutFlow_LHC} and \ref{tab:cutFlow_FCC}.

It is worth noting that the signal process may feature two jets with an invariant mass around the $W$ boson mass with possible further hadronic activity. We remark that in scenarios where the heavy neutrino mass is large its decay products can be strongly boosted, such that the hadronic decays of the $W$ bosons may be collimated, giving rise to a single jet instead of two.

\subsection{Standard Model Backgrounds}
\label{subsec:backgrounds}
The dominant SM backgrounds contributing to the $e^\pm \mu^\mp j j$ signature arise for instance from  the di-lepton final state with additional missing momentum due to processes with light neutrinos, or from the di-tau final state with both tau's decaying leptonically. In principle, these backgrounds can be rejected
with high signal efficiency by requiring the amount of missing energy in the final state to be small.
However, 
due to effects like the finite resolution of the missing momentum,
some backgrounds may still survive after such cuts. Thus, we expect that a full detector simulation, which is beyond the scope of the present analysis, can be important.

The background processes considered in our analysis are

\begin{enumerate}[label*=\arabic*.]

\item di-top in fully leptonical decays:
\begin{enumerate}[label*=\arabic*.]
\item $pp \to t\bar{t} \to (bW^+) (\bar{b}W^-) \to (b\, l^+ \nu) (\bar{b}\, l^- \bar{\nu}) $, where both $l$ can be either $e$ or $\mu$ ;
\end{enumerate}

\item di-boson with di-tau di-jet final states:
\begin{enumerate}[label*=\arabic*.]
\item $pp \to WZ \to (j j) (\tau^+ \tau^-)$;
\item $pp \to ZZ \to (j j) (\tau^+ \tau^-) $;
\end{enumerate}

\item tri-boson with at least 2 jets and at least 2 leptons (including taus):
\begin{enumerate}[label*=\arabic*.]
\item $pp \to WWZ \to (l\nu)(l\nu)(jj)$;
\item $pp \to WWZ \to (jj)(jj)(\tau^+ \tau^-)$;
\item $pp \to WWZ \to (jj)(l\nu)(\tau^+ \tau^-)$;
\item $pp \to WWZ \to (jj)(l\nu)(l^+ l^-)$.
\end{enumerate}

\end{enumerate}

For the tri-boson, if both taus decay leptonically, the final state will have 3 leptons. When one lepton is out of the detector range or mis-identified, it can still contribute to the backgrounds.
The other decay channels of $W$ or $Z$ bosons will either have no $e^\pm \mu^\mp$ final states or are lacking of jets. Therefore, they are not included in our analysis. 
The production cross sections corresponding to $t\bar{t}$, $WZ$, $ZZ$, $WWZ$ 
with decaying into the final states listed above
are about 3432 ($1.37\times10^5$), 1787 (5654), 468 (4483), 6.83 (95.5) fb at the HL-LHC (FCC-hh), respectively.

Furthermore, we checked many other possible background processes, including for instance all the processes listed above with one additional gluon jet or photon $(\gamma)$ in the final state, and also the processes 
$VVgg$
, $\gamma \mu\mu V$ with $V=Z,W$, and $\gamma \mu \nu W$.
We used an estimated rate of misidentifying $\gamma,g$ as an electron at FCC-hh of $\sim 10^{-3}$, comparable to the one at the LHC.
We found that especially the requirement of large transverse momenta of the $g,\gamma$, renders the cross sections of these processes much smaller 
than
the ones listed above, and we decided not to include them into our analysis.

\subsection{Simulation, Pre-selection and Analysis}
\label{subsec:simulation}
For the simulation of signal and background samples, we use MadGraph5 version 2.4.3~\cite{Alwall:2014hca} as the event generator. 
The parton shower and hadronization is done by Pythia6~\cite{Sjostrand:2006za}, while the detector simulations are completed by Delphes~\cite{deFavereau:2013fsa} with the ATLAS configuration card file (version 3.4.1) for the HL-LHC and with the FCC-hh configuration card file (October 2016 version) for the FCC-hh.

Based on the kinematics of the signal and backgrounds, in order to generate the events more effectively, we apply the following cuts at the 
simulation
level:
a minimal transverse momentum $p_T(j) >$ 20 GeV, $p_T(l) >$ 20 GeV and
the range of the pseudorapidity $|\eta(j)| <$ 10, $|\eta(l)| <$ 7 for jets (including b-jets) and leptons;
a maximal missing energy 
$\met <$ 30 GeV.
The cuts on $|\eta|$ do not affect the analysis because the detector geometry limits this range to be $|\eta| \lesssim\, 5$.
The cut on the missing energy are motivated from the prior knowledge that the signal does not produce missing energy at the parton level and only a limited amount can be created during reconstruction~\cite{Antusch:2016ejd}.
These cuts at the parton level enhance the quality of the background events and thus save the simulation time. 

The following pre-selection cuts are then applied on the simulation events:

\begin{enumerate}
\item Exactly 1 muon, exactly 1 electron, with opposite charges (i.e. $e^{\pm}\mu^{\mp}$ ); at least 2 jets; no b-jet and no taus;
\item Both jets and leptons with threshold cuts of $p_T > 30$ GeV;
\item Missing energy $\met < 20$ GeV.
\end{enumerate}

After the pre-selection cuts, the final state will have at least 2 light jets, 1 muon and 1 electron. The first two leading jets $j_1$ and $j_2$ are considered to be the jets from the final state W decay (see Fig.~\ref{fig:diagram}). To identify the lepton $l_N$ from the sterile neutrino decay, we combine the first two leading jets with each lepton and calculate the invariant masses corresponding to two combinations. The combination with invariant mass closer to the sterile neutrino mass indicates $l_N$, while the other lepton will be identified as the lepton $l_{W^*}$ from the off-shell $W^*$ decay.

Once the $l_{W^*}$ and $l_N$ are identified, the following 40 observables are input into the TMVA package~\cite{TMVA2007} to perform the Multi-Variate Analysis (MVA):

\begin{enumerate}[label*=\arabic*.]

\item global observables:
\begin{enumerate}[label*=\arabic*.]
\item the missing energy $\met$;
\item the scalar sum of the transverse momentum $p_T$ of all jets $H_T$;
\item the scalar sum of $p_T$ of all visible objects $p_T^{\rm vis}$.
\end{enumerate}

\item observables for the jets and leptons:
\begin{enumerate}[label*=\arabic*.]
\item $p_T$ and the pseudorapidity $\eta$ of the first two leading jets $j_1$ and $j_2$: $p_T(j_1)$, $\eta(j_1)$, $p_T(j_2)$, $\eta(j_2)$;
\item $p_T$, $\eta$ and the invariant mass $M$ of the system of $j_1$ and $j_2$: $p_T(j_1+j_2)$, $\eta(j_1+j_2)$, $M(j_1+j_2)$;
\item $p_T$ and $\eta$ of the lepton from the off-shell W decay $l_{W^*}$ and the lepton from the heavy neutrino N decay $l_N$: $p_T(l_{W^*})$, $\eta(l_{W^*})$, $p_T(l_N)$, $\eta(l_N)$;
\item $M$ of the system of $l_{W^*}$ and $l_N$: $M(l_{W^*}+l_N)$;
\item the pseudorapidity difference $\Delta\eta$ between jet and lepton: $\Delta\eta(j_1,l_{W^*})$, $\Delta\eta(j_2,l_{W^*})$, $\Delta\eta(j_1,l_N)$, $\Delta\eta(j_2,l_N)$;
\item the azimuthal angle difference $\Delta\phi$: $\Delta\phi(j_1,l_{W^*})$, $\Delta\phi(j_2,l_{W^*})$, $\Delta\phi(j_1,l_N)$, $\Delta\phi(j_2,l_N)$; 
\item the angular distance difference $\Delta R$: $\Delta R(j_1,l_{W^*})$, $\Delta R(j_2,l_{W^*})$, $\Delta R(j_1,l_N)$, $\Delta R(j_2,l_N)$.
\end{enumerate}

\item observables for the reconstructed N system:
\begin{enumerate}[label*=\arabic*.]
\item $p_T$, $\eta$, and $M$ of the system: $p_T(j_1+j_2+l_N)$, $\eta(j_1+j_2+l_N)$, $M(j_1+j_2+l_N)$;
\item $\Delta\eta$, $\Delta\phi$ and $\Delta R$ between the system of jets and $l_N$: $\Delta\eta(j_1+j_2, l_N)$, $\Delta\phi(j_1+j_2, l_N)$, $\Delta R(j_1+j_2, l_N)$.
\end{enumerate}

\item observables for the reconstructed off-shell $W^*$ system:
\begin{enumerate}[label*=\arabic*.]
\item $M$ of the system: $M(j_1+j_2+l_N+l_{W^*})$;
\item $p_T$, $\eta$, and $M$ of the system of jets and $l_{W^*}$: $p_T(j_1+j_2+l_{W^*})$, $\eta(j_1+j_2+l_{W^*})$, $M(j_1+j_2+l_{W^*})$;
\item $\Delta\eta$, $\Delta\phi$ and $\Delta R$ between the system of jets and $l_{W^*}$: $\Delta\eta(j_1+j_2, l_{W^*})$, $\Delta\phi(j_1+j_2, l_{W^*})$, $\Delta R(j_1+j_2, l_{W^*})$.
\end{enumerate}

\end{enumerate}

The details of the multivariate and statistical analysis are explained in the Appendix~\ref{app:mva}.

\section{Results}
\label{sec:results}

In this section, we present the analysis results for the HL-LHC and for the 100 TeV proton collider FCC-hh, which is also valid for the SppC with the same detector performance. We remind ourselves that the HL-LHC (FCC-hh) has center-of-mass energy $\sqrt{s}=$ 14 (100) TeV and that we consider a total integrated luminosity of 3 (20) $\iab$.

\subsection{Results at HL-LHC and FCC-hh}
\label{subsec:results}

To illustrate our results, we show the distributions of some selected observables after applying the pre-selection cuts for the signal with benchmark mass $M_N$ = 500 GeV 
(S, black with filled area), and SM backgrounds of $t\bar{t}$ (red), WZ (blue), ZZ (cyan), and WWZ (green) in Appendix~\ref{app:distributions}.
The Fig.~\ref{fig:HLLHC_distributions} and Fig.~\ref{fig:FCChh_distributions} are for the HL-LHC and FCC-hh, respectively.

\begin{figure}[htbp]
\centering
\includegraphics[width=7.5cm,height=5cm]{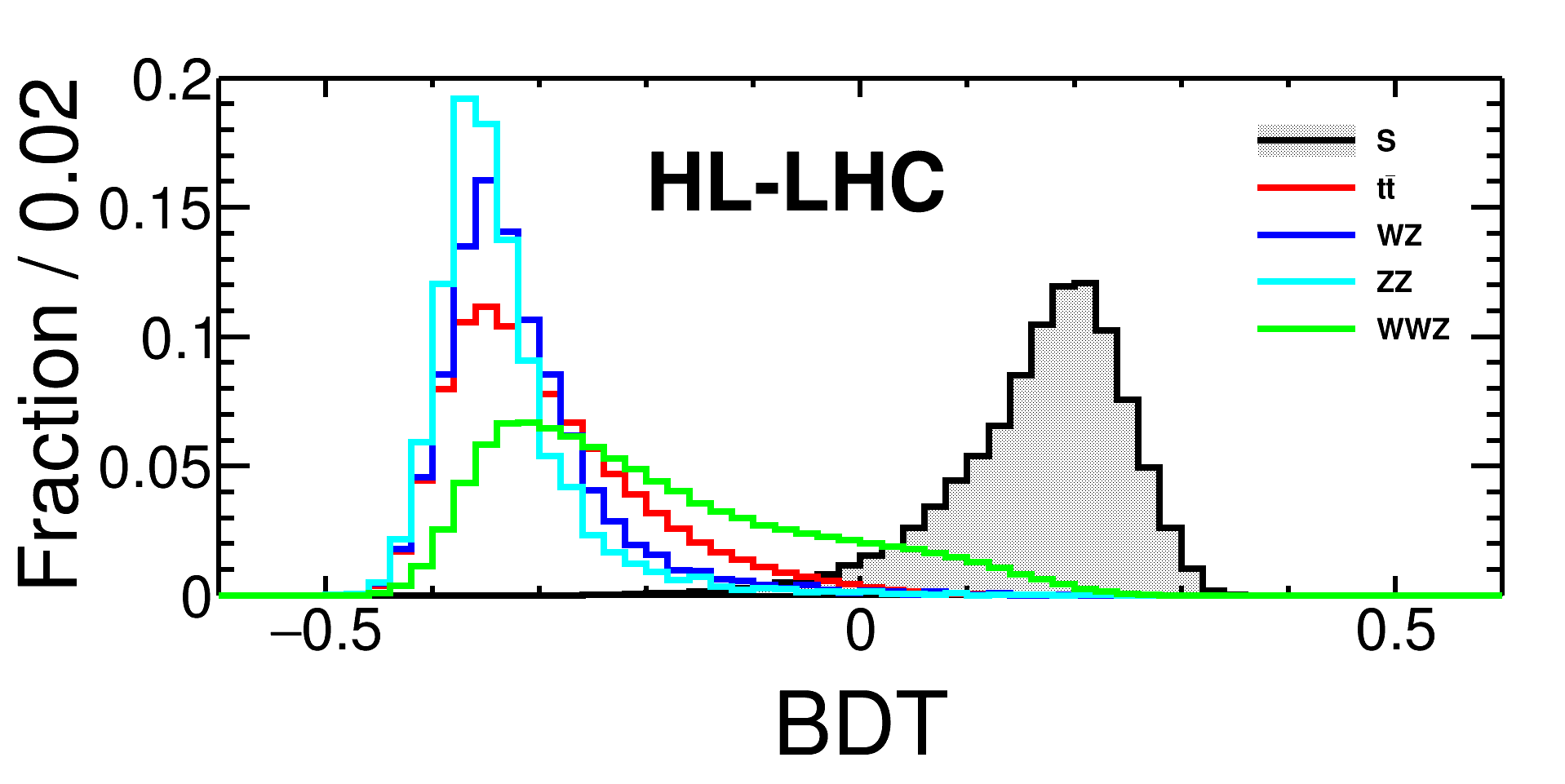}
\includegraphics[width=7.5cm,height=5cm]{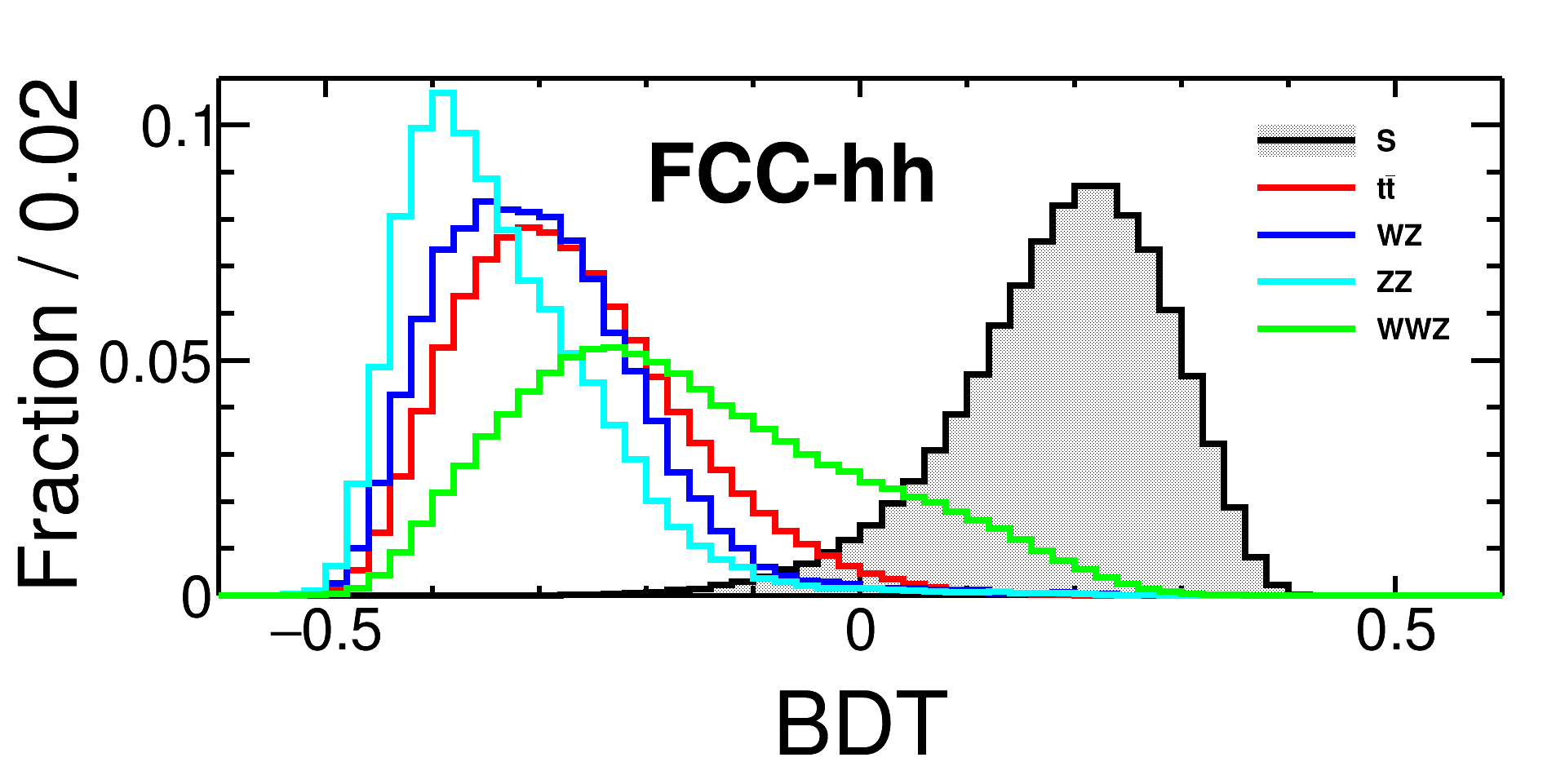}
\caption{Distributions of BDT response at the 
HL-LHC (left) and FCC-hh (right)
for signal with $M_N$ = 500 GeV (S, black with filled area), and for SM backgrounds including $t\bar{t}$ (red), WZ (blue), ZZ (cyan),
and WWZ (green).}
\label{fig:BDT}
\end{figure}

One can see from Fig.~\ref{fig:HLLHC_distributions} that the distributions of signal and SM background are very different for the given examples. For brevity, we describe here only a few of them at the HL-LHC:
The distributions of the $M(j_1+j_2+l_N)$ of the signal peaks sharply around the sterile neutrino mass 500 GeV, while all backgrounds peak below 250 GeV;
In the di-jet invariant mass $M(j_1+j_2)$ plot, the signal and WZ peaks around the W boson mass, while ZZ and WWZ peak around the Z boson mass and $t\bar{t}$ has a flat peak around 110 GeV;
In the di-lepton invariant mass $M(l_{W^*}+l_N)$ plot, the backgrounds WZ and ZZ peak around 70 GeV, and $t\bar{t}$ and WWZ peak around 100 GeV, while the signal has a very flat peak around 400 GeV;
For the distributions of $M(j_1+j_2+l_N+l_{W^*})$, $p_T(l_N)$, $p_T(j_1+j_2)$ and $p_T^{\rm vis}$, the signal peaks at larger values compared to the backgrounds.
Other useful distributions to distinguish signal from background exist, for instance
$\met$ and angular observables, 
which
we list in section~\ref{subsec:simulation}.

\begin{table}[htbp]
\begin{center}
\scalebox{0.75}{
\begin{tabular}{c|c|cccccc|cccc}
\hline
\hline
\multicolumn{2}{c|}{\multirow{2}{*}{Cuts}} & \multicolumn{6}{c|}{$M_N$ [GeV]} & \multicolumn{4}{c}{Background} \\
\multicolumn{2}{c|}{ } & 200 & 400 & 500 & 600 & 800 & 1000 & $t\bar{t}$ & WZ & ZZ & WWZ \\
\hline
\multicolumn{2}{c|}{initial} & $1.56\times10^4$ & 1307 & 563 & 275 & 83.2 & 30.7 & $1.03\times10^7$ & $5.36\times10^6$ & $1.40\times10^6$ & $2.05\times10^4$   \\
\hline
\multirow{3}{*}{pre-sel.}
 & cut 1 & 2545 & 260 & 109 & 50.6 & 14.1 & 5.0 & $3.26\times10^5$ & $2.63\times10^4$ & 6008 & 343   \\
 & cut 2 & 1830 & 229 & 97.7 & 45.2 &12.4 & 4.4 & $1.83\times10^5$ & 1462 & 337 & 164   \\
 & cut 3 & 1376 & 130 & 46.9 & 18.5 & 3.7 & 0.99 & $5.44\times10^4$ &  265 & 64 & 58   \\
\hline
\multirow{6}{*}{BDT} 
 & $>0.2013$ & 111 &    - &    - &   - &    - &  -   & 19.1    & 0.10 & 0.027 & 0.56     \\
 & $>0.2162$ &   - & 37.8 &    - &   - &    - & -    & 2.3     & -    & 0.027 & 0.41     \\
 & $>0.2148$ &   - &    - & 13.9 &   - &    - & -    & 0.63    & -    & 0.014 & 0.16     \\
 & $>0.2263$ &   - &    - &    - & 3.6 &    - &  -   & 0.13    & -    & 0.014 & 0.046    \\
 & $>0.2264$ &   - &    - &    - &   - & 0.63 &   -  & 0.0068  & -    &     - & 0.013    \\
 & $>0.2348$ &   - &    - &    - &   - &    - & 0.15 & 0.00012 & -    &     - & 0.0041   \\
\hline
\hline
\end{tabular}
}
\end{center}
\caption{Numbers of events at each cut stage for signals with fixed $|V_{ lN}|^2=10^{-2}$ and different sterile neutrino masses $M_N$ and for background processes. The numbers correspond to an integrated luminosity of $3~\mathrm{ab}^{-1}$ at the HL-LHC.}
\label{tab:cutFlow_LHC}
\end{table}

As described in section \ref{subsec:simulation}, all the 40 observables 
listed in that section are input into the TMVA.
We utilize the Boosted Decision Trees (BDT) method to perform the multivariate analysis.
The distributions of the BDT response for the signal with $M_N$ = 500 GeV 
(S, black with filled area), and for the SM backgrounds including $t\bar{t}$ (red), WZ (blue), ZZ (cyan),
and WWZ (green) are shown in Fig.~\ref{fig:BDT} for the HL-LHC (left) and the FCC-hh (right). The BDT response shows that a very good separation between the signal and background is possible. For WWZ background process, although it has larger mixing with the signal, due to its relatively small initial production cross section, its final contributions to the backgrounds after the optimized BDT cut are still limited, cf.\ Table~\ref{tab:cutFlow_LHC} and Table~\ref{tab:cutFlow_FCC}.

\begin{table*}[htbp]
\begin{center}
\scalebox{0.6}{
\begin{tabular}{c|c|cccccc|cccc}
\hline
\hline
\multicolumn{2}{c|}{\multirow{2}{*}{Cuts}} & \multicolumn{6}{c|}{$M_N$ [GeV]} & \multicolumn{4}{c}{Background}    \\
\multicolumn{2}{c|}{ } & 200 & 400 & 500 & 600 & 800 & 1000 & $t\bar{t}$ & WZ & ZZ & WWZ   \\
\hline
\multicolumn{2}{c|}{initial} & $1.78\times10^6$ & $2.14\times10^5$ & $1.07\times10^5$ & $6.03\times10^4$ & $2.38\times10^4$ & $1.13\times10^4$ & $2.75\times10^9$ & $1.13\times10^8$ & $8.97\times10^7$ & $1.91\times10^6$   \\
\hline
\multirow{3}{*}{pre-sel.}
 & cut 1 & $3.84\times10^5$ & $5.98\times10^4$ & $3.03\times10^4$ & $1.70\times10^4$ & 6347 & 2856 & $6.08\times10^7$ & $1.96\times10^6$ & $1.46\times10^6$ & $5.45\times10^4$   \\
 & cut 2 & $3.39\times10^5$ & $5.76\times10^4$ & $2.95\times10^4$ & $1.66\times10^4$ & 6257 & 2824 & $3.61\times10^7$ & $6.20\times10^4$ & $4.24\times10^4$ & $1.96\times10^4$   \\
 & cut 3 & $2.90\times10^5$ & $4.36\times10^4$ & $2.10\times10^4$ & $1.12\times10^4$ & 3722 & 1484 & $9.08\times10^6$ & 7090 & 5497 & 6657   \\
\hline
\multirow{6}{*}{BDT} 
 & $>0.2935$ & 6611 & - & - & - & - & - & 238.4 & 0.6 & 0.5 & 15.9   \\
 & $>0.2827$ & - & 5762 & - & - & - & - &  81.5 & 0.9 & 0.7 & 20.3   \\
 & $>0.2654$ & - & - & 4666 & - & - & -  &  53.8 & 0.3 & 0.5 & 16.4   \\
 & $>0.2611$ & - & - & - & 2701 & - & -  &  33.9 &   - &   - &  8.9   \\
 & $>0.2428$ & - & - & - & - & 1261 & -   &  27.1 & 0.3 &   - &  6.7   \\
 & $>0.2262$ & - & - & - & - & - & 693   &  27.6 & 0.3 &   - &  6.7   \\
\hline
\hline
\end{tabular}
}
\end{center}
\caption{
Numbers of events at each cut stage for signals with fixed $|V_{ lN}|^2=10^{-2}$ and different sterile neutrino masses $M_N$ and for background processes.
The numbers correspond to an integrated luminosity of $20~\mathrm{ab}^{-1}$ at the FCC-hh.
}
\label{tab:cutFlow_FCC}
\end{table*}

In Table~\ref{tab:cutFlow_LHC}, we show the numbers of events at each cut stage for signals with $|V_{ lN}|^2=10^{-2}$ and different sterile neutrino masses $M_N$ and for background processes of $t\bar{t}$, WZ, ZZ, and WWZ at the HL-LHC with $3~\mathrm{ab}^{-1}$ integrated luminosity.
The numbers of events at the FCC-hh with $20~\mathrm{ab}^{-1}$ integrated luminosity are presented in Table~\ref{tab:cutFlow_FCC}.
Since the kinematical distributions vary with $M_N$,
the BDT cuts are optimized for different masses.

\begin{figure}[htbp]
\centering
\includegraphics[width=7.5cm,height=5cm]{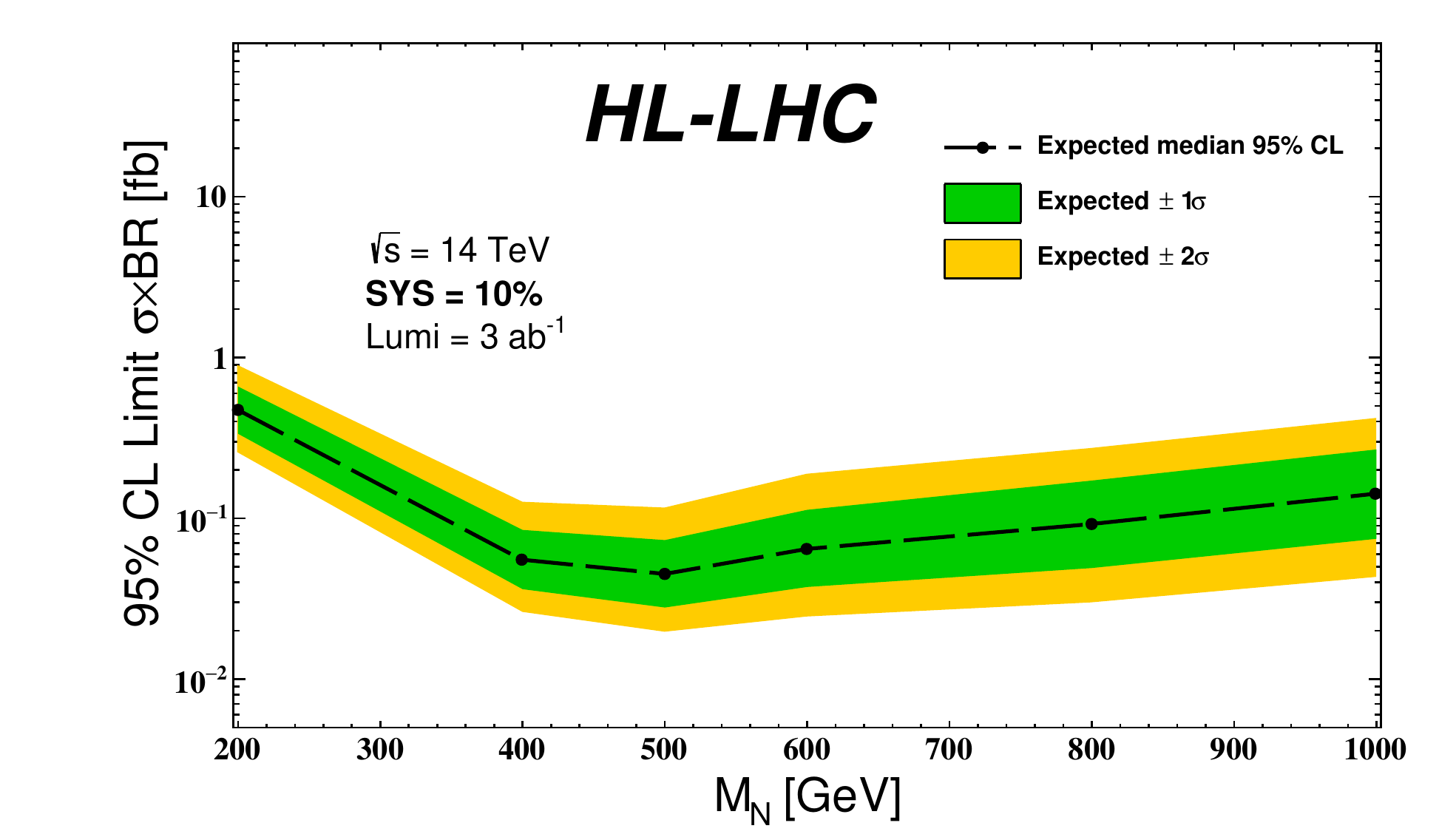}
\includegraphics[width=7.5cm,height=5cm]{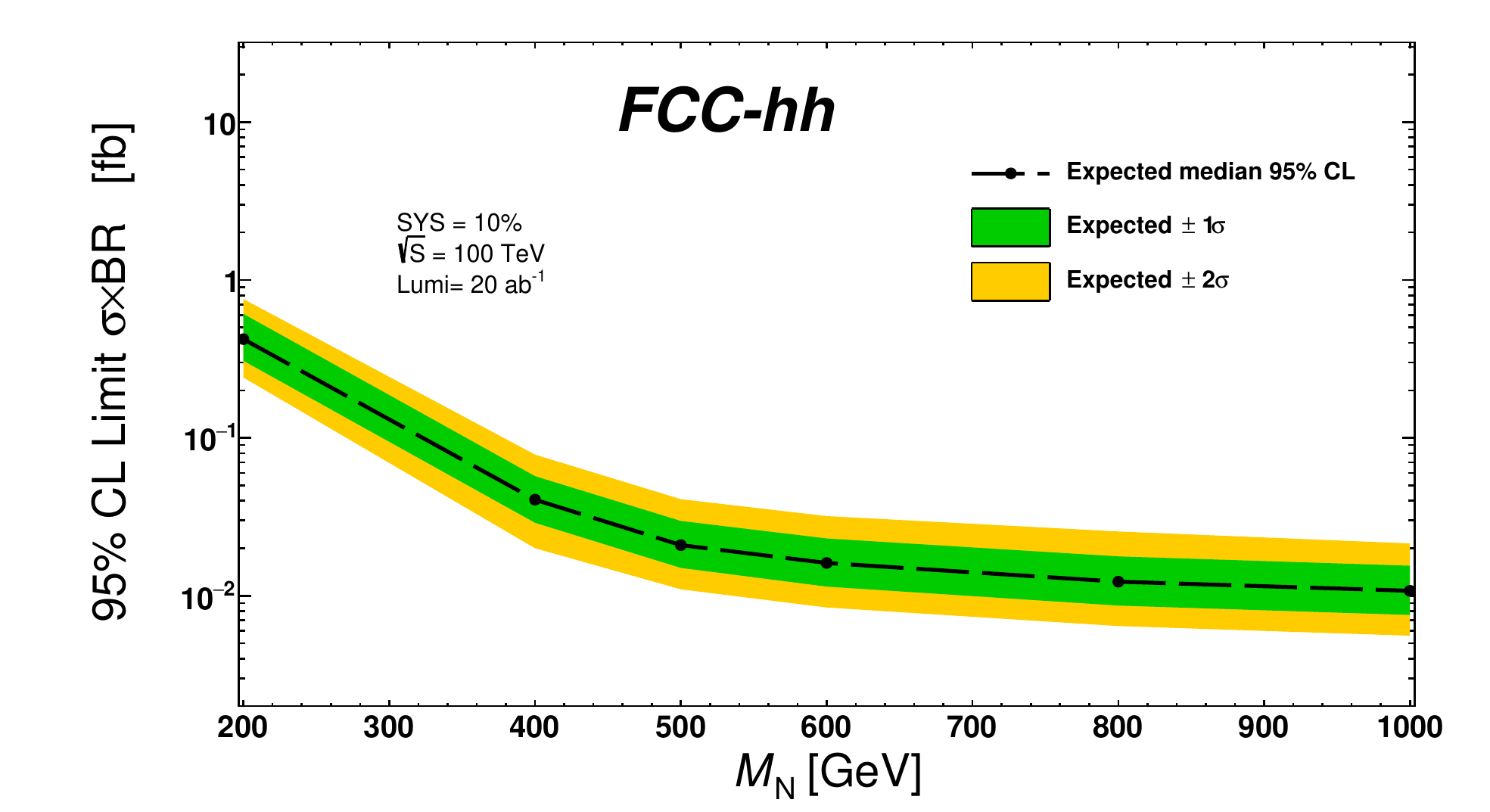}
\caption{Expected 
limits
on  the production cross section times branching ratio 
$\sigma(p p \to l^\pm N)\times {\rm BR} (N \to l^{\mp} jj)$
in fb when testing the signal hypothesis at the HL-LHC (left) with $\sqrt{s}=14$ TeV and 3 $\iab$ and at the FCC-hh (right) with $\sqrt{s}=100$ TeV and 20 $\iab$, including the 1 and 2-$\sigma$ confidence interval.
These limits have 
been derived based on the analysis of the $e^\pm \mu^\mp jj$ final state.}
\label{fig:sigma_limit}
\end{figure}

Based on our analysis, the prospects for sterile neutrino searches via the opposite sign mixed-flavor dilepton plus di-jet (i.e. $e^\pm \mu^\mp jj$) 
including a systematic uncertainty of $\delta_{\rm sys}=10\%$ 
on the background
are derived,
using the Higgs Analysis-Combined Limit tool~\cite{combine}, for details see the explanations in the Appendix~\ref{app:mva}.
In Fig.~\ref{fig:sigma_limit}, we show the expected limit on 
the production cross section times branching ratio 
$\sigma(p p \to l^\pm N)\times {\rm BR} (N \to l^{\mp} jj)$
in fb when
testing the signal hypothesis at the HL-LHC 
(left)
with $\sqrt{s}=14$ TeV and 3 $\iab$ and at the FCC-hh 
(right)
with $\sqrt{s}=100$ TeV and 20 $\iab$, including the 1 and 2-$\sigma$ confidence interval.
The figure shows that the total production cross section for this final state can be tested at the HL-LHC and FCC-hh for values of ${\cal O}(0.1)$ and ${\cal O}(0.01)$ fb, respectively.
It is worthy of note that the decline of the production cross section for increasing masses is (at least partially) compensated for by the increased BDT efficiency, such that the limits on the total cross section remain more or less flat.

\begin{figure}[htbp]
\centering
\includegraphics[width=7.5cm,height=5cm]{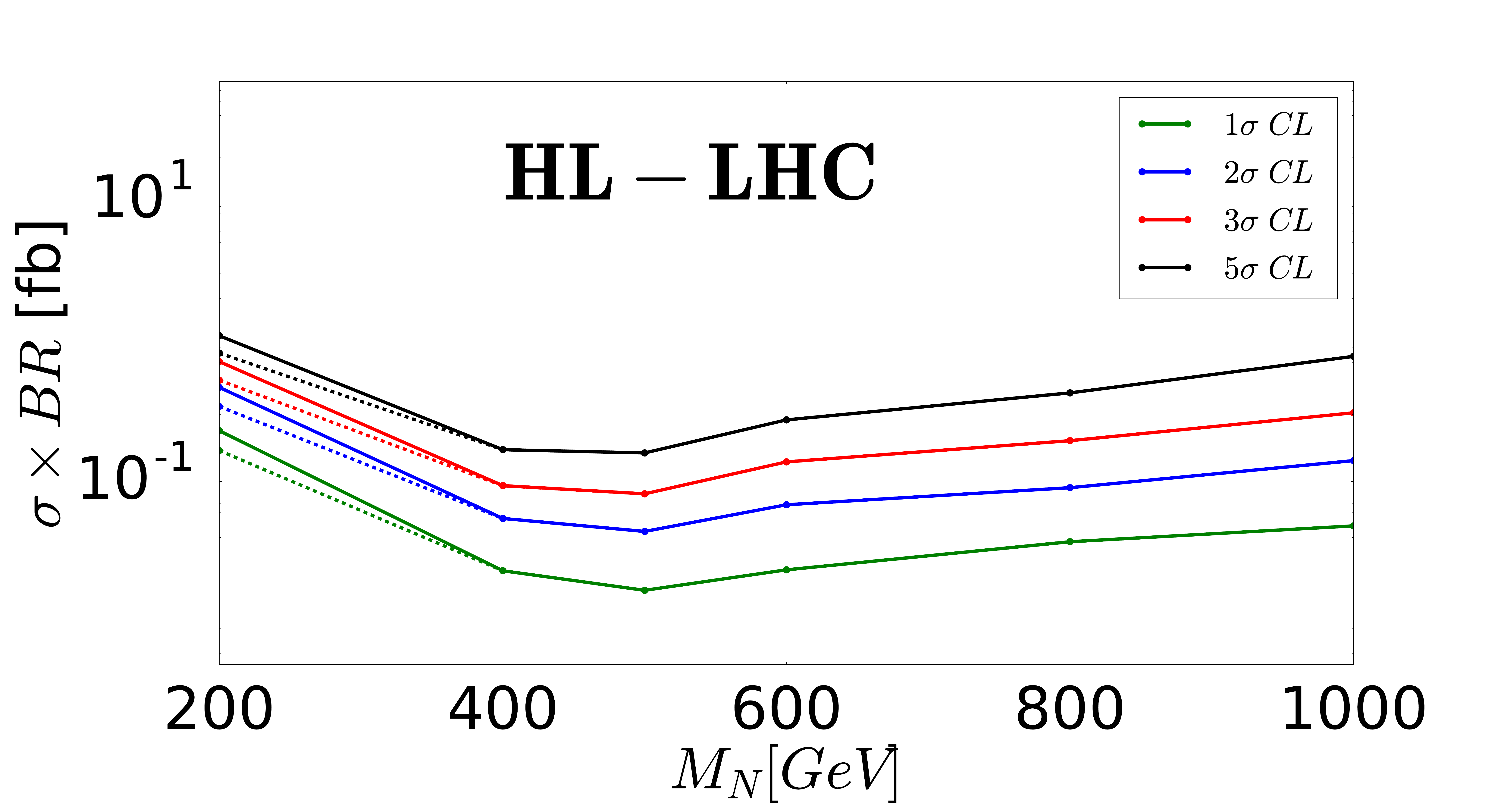}
\includegraphics[width=7.5cm,height=5cm]{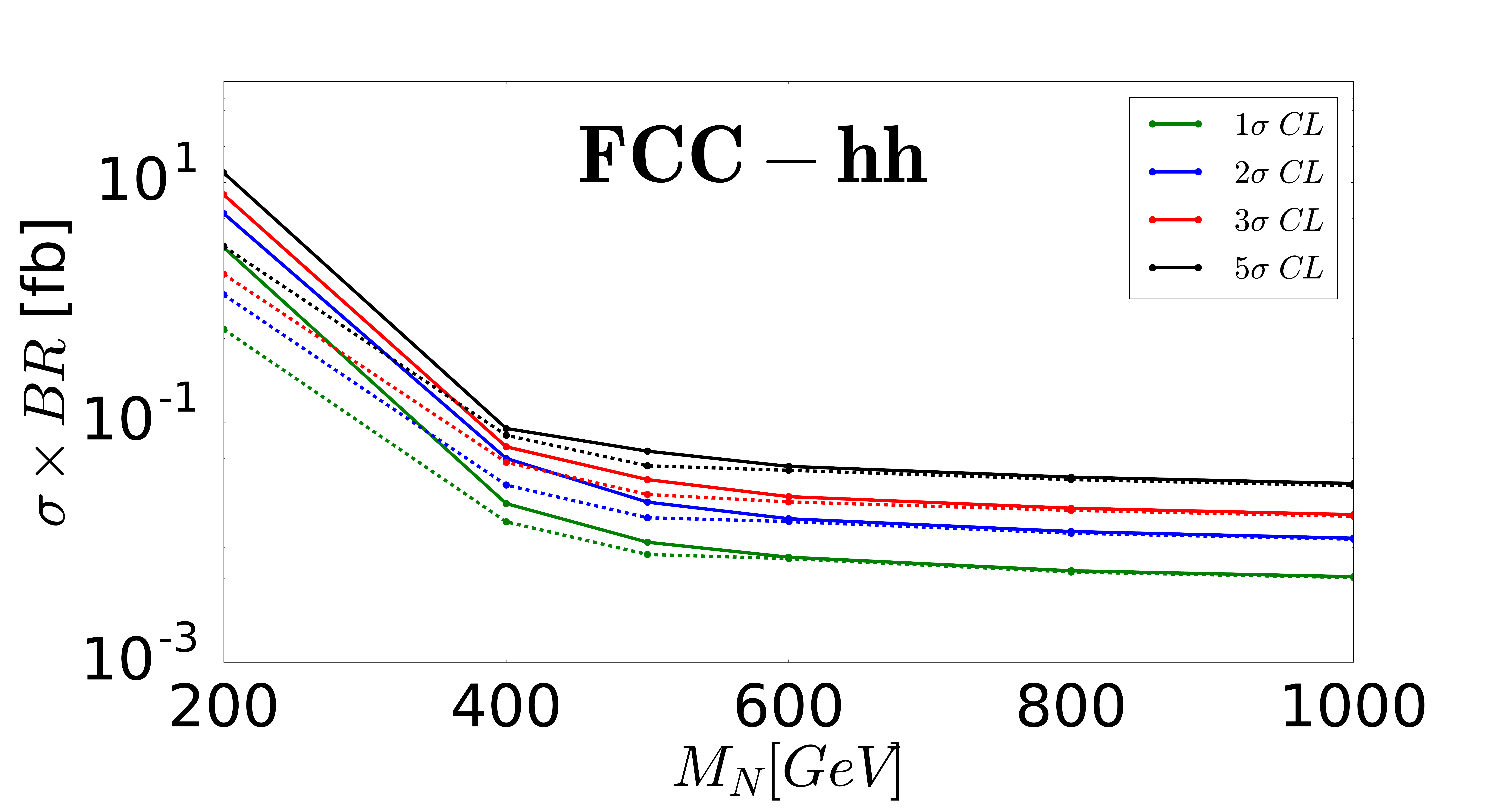}
\caption{
Same as
Fig.\ \ref{fig:sigma_limit}, including the 1, 2, 3 and 5-$\sigma$ 
median expected
limits on the production cross section times branching ratio 
$\sigma(p p \to l^\pm N)\times {\rm BR} (N \to l^{\mp} jj)$
in fb at the HL-LHC (left) with 3 $\iab$ luminosity and at the FCC-hh (right) with $\sqrt{s}=100$ TeV and 20 $\iab$ luminosity. 
In both panels the solid (dashed) line denotes that a 10\% (0\%) systematic uncertainty on the background is considered.}
\label{fig:sigma_limit_median}
\end{figure}

In Fig.~\ref{fig:sigma_limit_median}, we show the 1, 2, 3 and 5-$\sigma$ median expected limits on the production cross section times branching ratio
$\sigma(p p \to l^\pm N)\times {\rm BR} (N \to l^{\mp} jj)$
in fb at the HL-LHC (left) with $\sqrt{s}=14$ TeV and 3 $\iab$ luminosity and at the FCC-hh (right) with $\sqrt{s}=100$ TeV and 20 $\iab$ luminosity. 
In this figure, the solid (dashed) line denotes that 10\% (0\%) systematic uncertainty on the background is considered.
Comparing the solid and dashed curves, one can see that as sterile neutrino mass $M_N$ decreases, the effects of the systematic uncertainty on the background become more obvious.
This is 
because
that the number of background events after the BDT cut will increase as $M_N$ decreases (cf. Table~\ref{tab:cutFlow_LHC} and Table~\ref{tab:cutFlow_FCC}).
When $M_N = 500$ GeV, with 0\% systematic uncertainty on background, the 2 (5)-$\sigma$ limit on the $\sigma \times {\rm BR}$ is $4.4\times 10^{-2}$($1.5\times 10^{-1}$) fb at the HL-LHC, while it is $1.6\times 10^{-2}$($4.3\times 10^{-2}$) fb at the FCC-hh.

\begin{figure}[htbp]
\centering
\includegraphics[width=7.5cm,height=5cm]{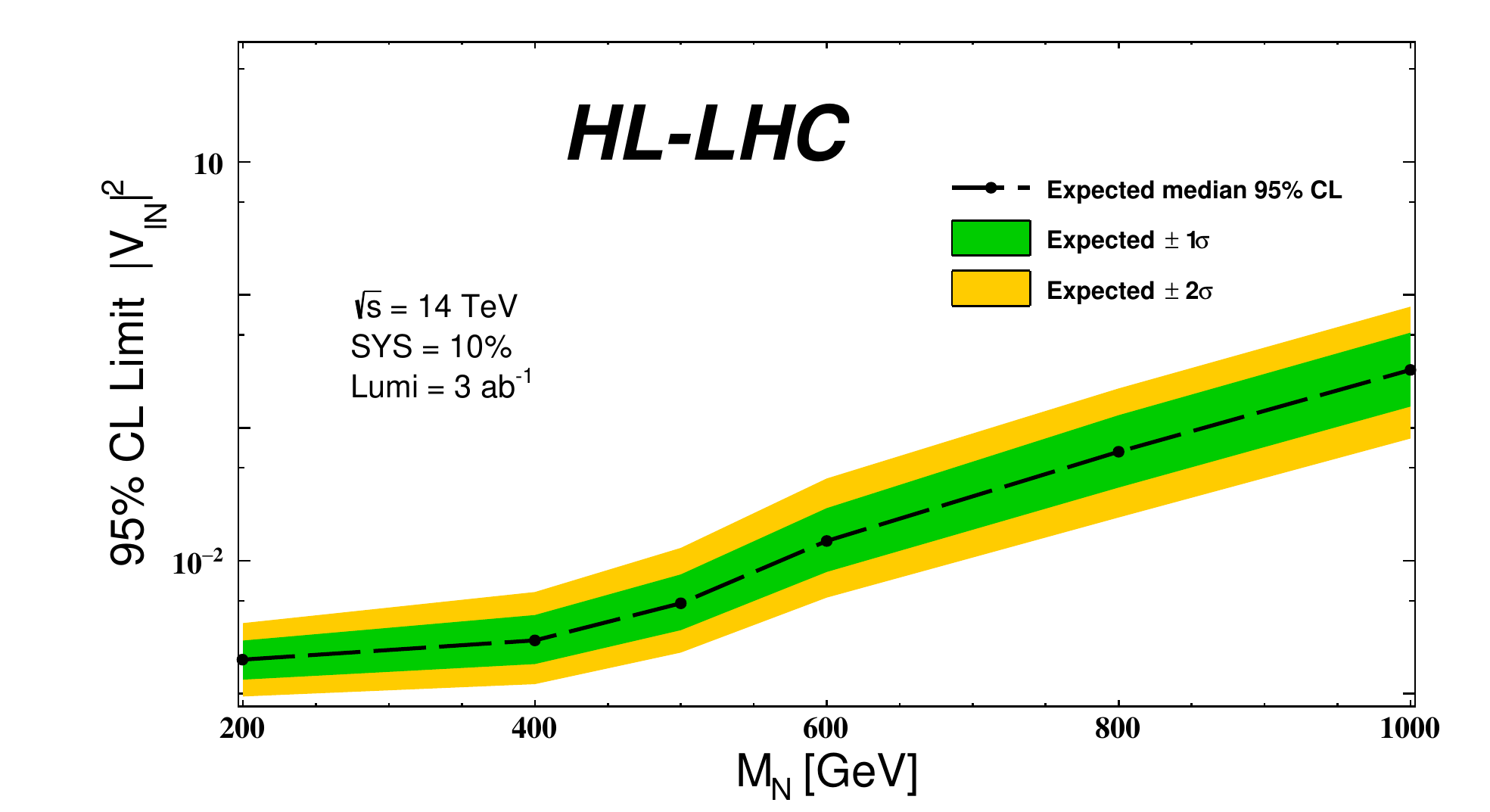}
\includegraphics[width=7.5cm,height=5cm]{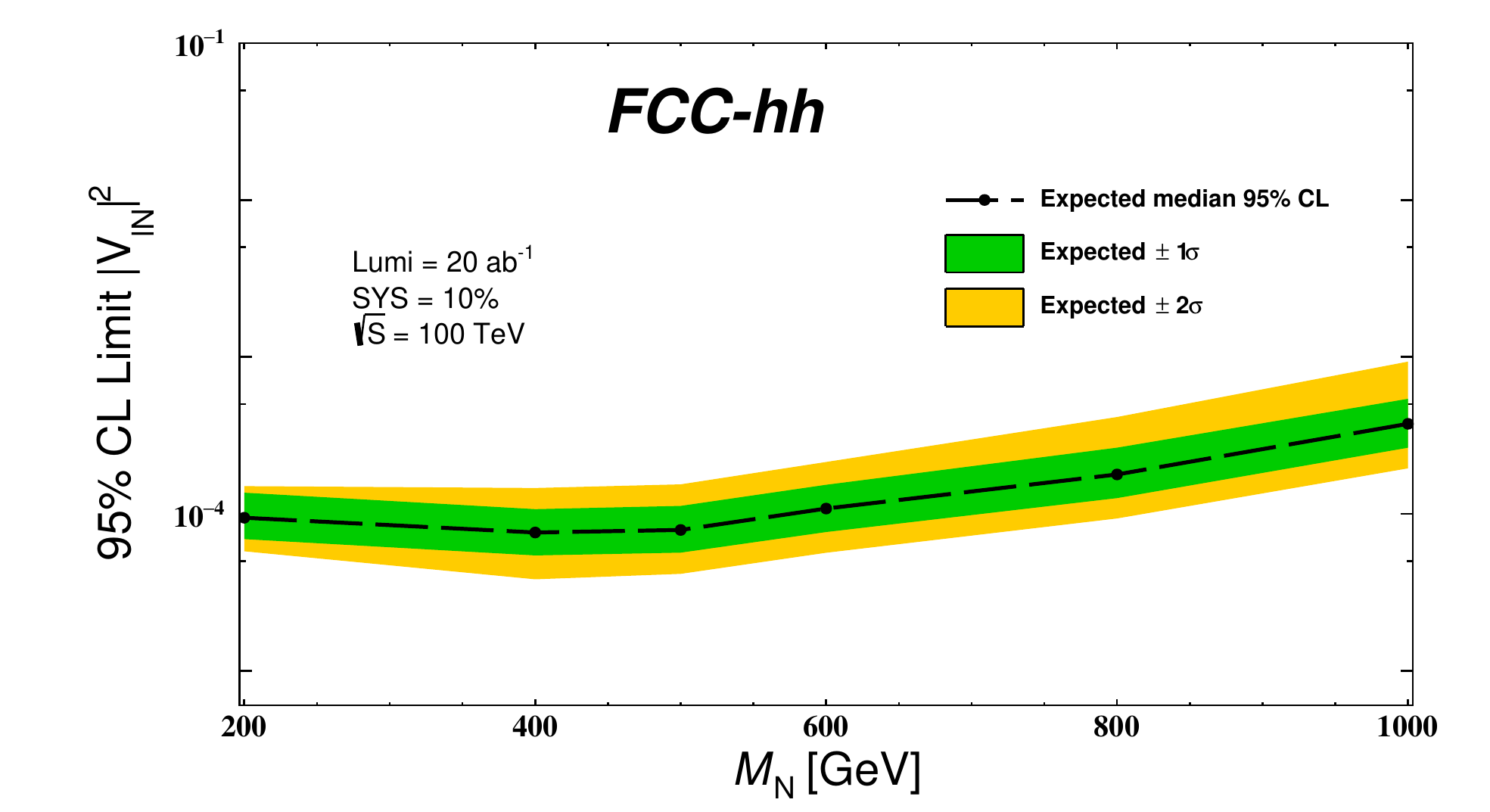}
\caption{
Expected 
limits
 on the parameter $|V_{ lN}|^2$ when testing the signal hypothesis for $|V_{ lN}|^2=|V_{ eN}|^2=|V_{ \mu N}|^2$ and $|V_{ \tau N}|^2=0$, including the 1 and 2-$\sigma$ confidence interval.
The left (right) panel denotes the limit for the HL-LHC (FCC-hh) with $\sqrt{s}=$ 14 (100) TeV and 3 (20) $\iab$ luminosity. 
These limits have
been derived based on the analysis of the $e^\pm \mu^\mp jj$ final state.
}
\label{fig:VlN_limit}
\end{figure}

Using the assumption in Eq.~\eqref{eqn:assumption} for the active-sterile mixing angles, we can
convert the limits 
from Fig.\ \ref{fig:sigma_limit_median} into limits
on $|V_{ lN}|^2$, cf.\ the definition in Eq.\ \eqref{eq:V_lN}.
We show the resulting expected median limit on the total active-sterile mixing $|V_{ lN}|^2$ in Fig.~\ref{fig:VlN_limit} for the HL-LHC (left) with $\sqrt{s}=14$ TeV and 3 $\iab$ and at the FCC-hh (right) with $\sqrt{s}=100$ TeV and 20 $\iab$, including the 1 and 2$\sigma$ confidence interval.
When extracting these limits, a systematic uncertainty of 10\% on the background has been considered.
It is worthwhile to point out that these results are quantitatively close to the first estimates from Ref.\ \cite{Antusch:2016ejd}.

\begin{figure}[htbp]
\centering
\includegraphics[width=7.5cm,height=5cm]{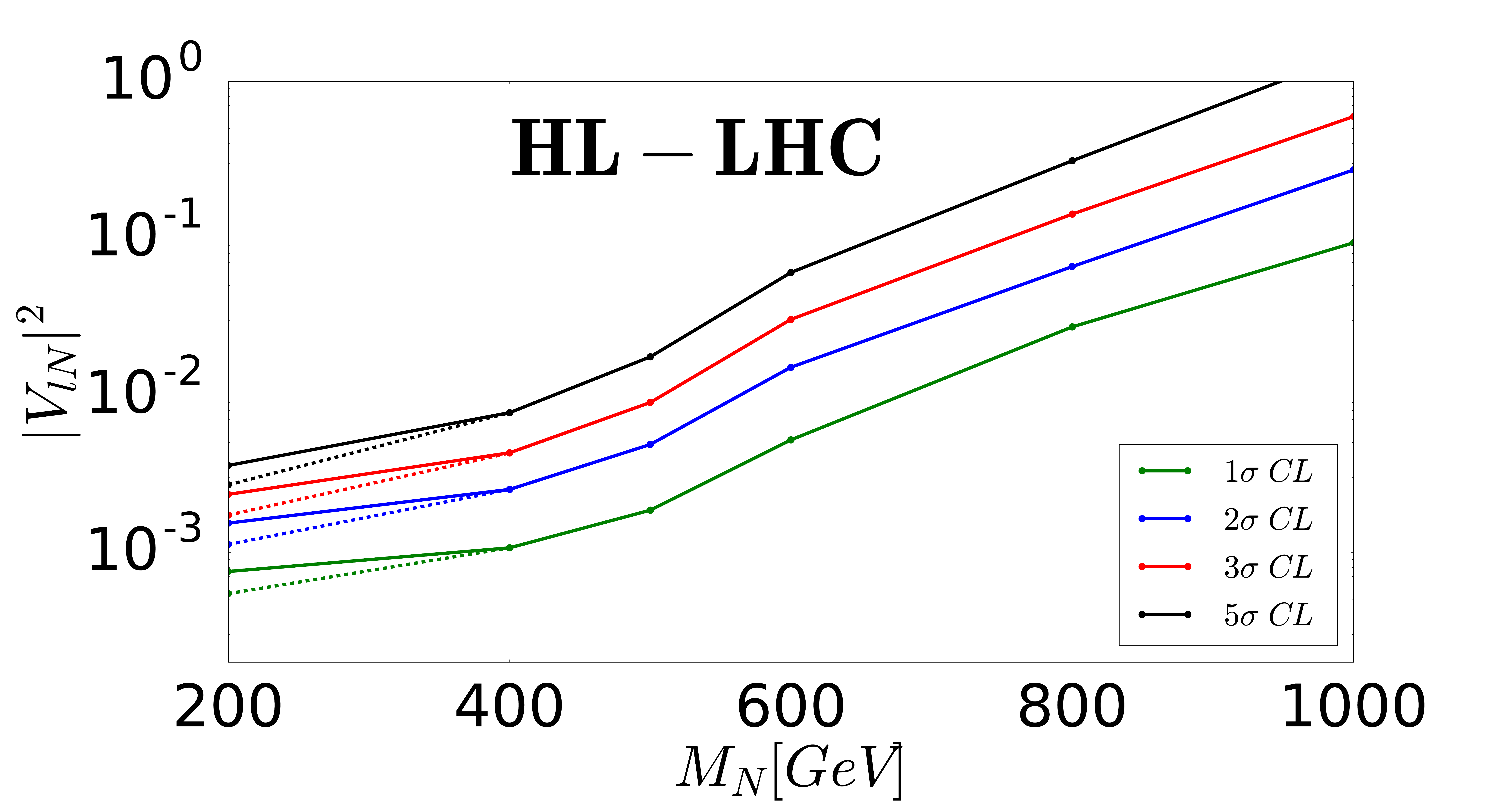}
\includegraphics[width=7.5cm,height=5cm]{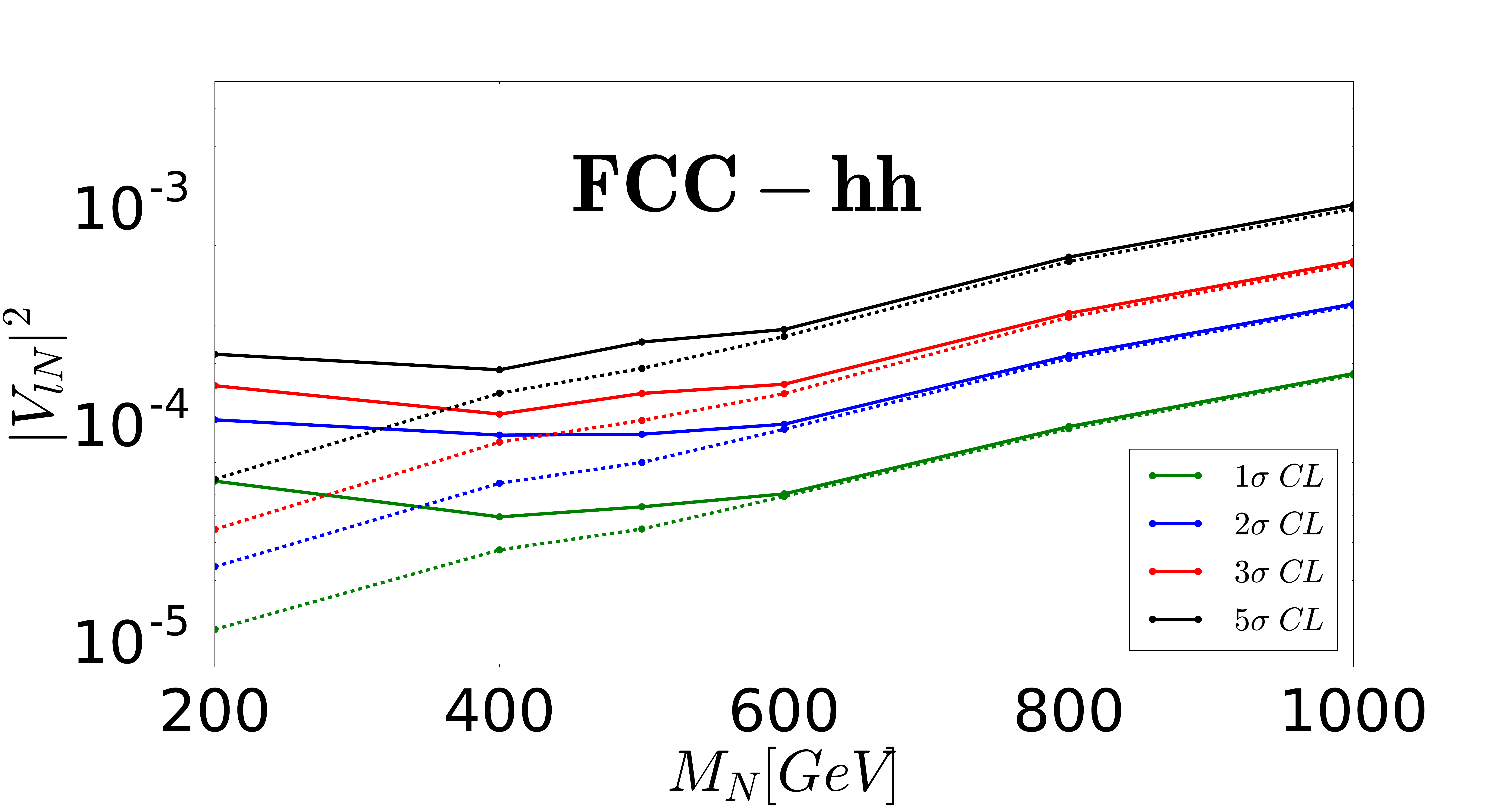}
\caption{
Same as Fig. \ref{fig:VlN_limit}, including the 1, 2, 3 and 
5-$\sigma$ median expected limits on the parameter $|V_{ lN}|^2$ for $|V_{ lN}|^2=|V_{ eN}|^2=|V_{ \mu N}|^2$ and $|V_{ \tau N}|^2=0$,
at the HL-LHC (left) with 3 $\iab$ luminosity and at the FCC-hh (right) with $\sqrt{s}=100$ TeV and 20 $\iab$ luminosity. 
In both panels the solid (dashed) line denotes that a 10\% (0\%) systematic uncertainty on the background is considered.
}
\label{fig:VlN_limit_median}
\end{figure}

In Fig.~\ref{fig:VlN_limit_median}, we show the 1, 2, 3 and 
5-$\sigma$
median expected limits on the total active-sterile mixing squared $|V_{ lN}|^2$  for 
the HL-LHC (left panel) and the FCC-hh (right panel),
including a systematic uncertainty of 0\% (dashed) and 10\% (solid) on the background.
Comparing the solid and dashed curves, one can see that
at the HL-LHC (FCC-hh), when $M_N < $ 400 (600) GeV, the effects of 10\% systematic uncertainty on the background become visible. For 200 GeV mass point, due to much larger background events after the BDT cut, the systematic uncertainty can weaken the limits greatly. Therefore, to enhance the discovery power for sterile neutrino with small masses, controlling the systematic uncertainty at such future colliders will be very important.
When $M_N = 500$ GeV, with 0\% systematic uncertainty on the background, the 2 (5)-$\sigma$ limit on the $|V_{ lN}|^2$ are $4.9\times 10^{-3}$($1.7\times 10^{-2}$) at the HL-LHC, while it is $7.0\times 10^{-5}$($1.9\times 10^{-4}$) at the FCC-hh.

\subsection{Discussion}
\label{subsec:discussion}
We note that 
our
results on the sensitivity of the proton-proton colliders are qualitatively identical to those in Ref.~\cite{Antusch:2016ejd}.
Moreover, the sensitivity is comparable to the analyses that consider lepton-number violating final states, cf.\ e.g.\ \cite{Deppisch:2015qwa}.

An important low energy constraint exists, that also tests the here considered active-sterile mixings: the $\mu\to e\gamma$ measurement from the 
Mu to E Gamma (MEG)
collaboration. Via the null result in their searches for the process $\mu \to e \gamma$ \cite{Mori:2016vwi} they put stringent limits on the combination
$|V_{ e N} V_{ \mu N}|$
(which is equal to $|\theta_e^* \theta_\mu|$) \cite{Antusch:2014woa,Antusch:2015mia}.
Indeed, finding a signal in the $e^\pm \mu^\mp jj$ channel at the HL-LHC or the FCC would be in tension with the present constraints from MEG.

It is interesting to consider the LFV dilepton dijet signature with one tau lepton in the final state. 
The relevant active-sterile mixing 
parameters that are tested in this way are then $|V_{eN} V_{\tau N}|$ and $|V_{\mu N} V_{\tau N}|$,
respectively. The present constraints on these mixing angle combinations are much weaker compared to those derived from the MEG result, cf.\ e.g.\ Ref.\ \cite{Antusch:2014woa} and references therein. This could mean a great discovery potential in these channels, if our results for the sensitivities would also hold, at least approximately, for the $e^\pm \tau^\mp jj$ and $\mu^\pm \tau^\mp jj$ final states.

However, including the tau flavor necessitates to reconstruct the tau lepton either from a muon or an electron, which requires the finding of a non-vanishing impact parameter and inserts additional missing momentum from the neutrino associated to a tau decay. More promising is the reconstruction of a tau from its hadronic decays, which, on the other hand, introduces many additional backgrounds involving heavy quarks.

Heavy neutrinos with 
masses above
1 TeV are produced dominantly via the $W\gamma$ fusion processes.
The kinematics of the final state 
particles
is very similar to the ones studied in our analysis.
We therefore assume, that for $M>1$ TeV, the sensitivity via $W\gamma$ fusion becomes better compared to our results, such that the latter comprise a conservative limit on these parameters.

\section{Conclusions}
\label{sec:summary}
Low scale seesaw scenarios allow for large active-sterile neutrino mixings and heavy neutrinos with masses that are kinematically accessible at particle colliders.
Due to the approximate symmetry,
lepton number violation is suppressed in these scenarios,
which motivates the study of lepton flavour violating (LFV) but lepton number conserving (LNC) signal channels.

In this article we investigated 
the most promising sterile neutrino signature of this type, based on parton level studies from previous works, the LFV but LNC final state  $e^\pm \mu^\mp jj $. 
This final state does not have SM backgrounds at the parton level, such that the signal and backgrounds can be well separated via a thorough analysis of the distributions from a number of kinematic observables.

For the active-sterile neutrino mixings we assumed, for definiteness, $|V_{ l N}|^2=|\theta_e|^2=|\theta_\mu|^2$ and $|V_{ \tau N}|^2 = |\theta_\tau|^2=0$. 
We remark that the $e^\pm \mu^\mp jj $ 
signature
is sensitive to
$|V_{ eN}V_{ \mu N}|^2/\sum_{\alpha} |V_{ \alpha N}|^2$,
which means that it is suppressed if one of the two active-sterile mixing 
parameters
is much larger than the other ones, while the signal rate is maximal for the case we assumed 
(for fixed $ \sum_\alpha |V_{\alpha N}|^2$).

We considered the HL-LHC (FCC-hh/SppC) with 
$\sqrt{s}=$ 14 (100) TeV and a total integrated luminosity of 3 (20) $\iab$. 
We simulated large event samples for the signal and for the dominant SM backgrounds processes 
(di-top, di-boson, and tri-boson) including parton shower, hadronization and fast detector simulation.
Forty kinematic observables are constructed from each event and are fed into a multivariate analysis tool to perform a BDT analysis. We derive the 1, 2, 3, and 5-$\sigma$ limits on the production cross section times branching ratio $\sigma(p p \to l^\pm N)\times {\rm BR} (N \to l^{ \mp} jj)$,
and recast it as a limit on the active-sterile mixing parameter $|V_{ lN}|^2$.
The result is comparable to the previous estimates obtained in Ref.\ \cite{Antusch:2016ejd}, but more robust.

We find that, under our assumptions and
for the benchmark $M_N=500$ GeV, when ignoring systematic uncertainties at the HL-LHC and the FCC-hh/SppC, the resulting 2 (5)-$\sigma$ sensitivities on $|V_{ l N}|^2$ are $4.9\times 10^{-3}$ ($1.7\times 10^{-2}$) and $7.0\times 10^{-5}$ ($1.9\times 10^{-4}$), while the 2 (5)-$\sigma$ limits on the production cross section times branching ratio $\sigma \times {\rm BR}$ are $4.4\times 10^{-2}$($1.5\times 10^{-1}$) fb and $1.6\times 10^{-2}$ ($4.3\times 10^{-2}$) fb, respectively.
At the FCC-hh, the reduced production rate for larger masses is partially compensated by the signal efficiency, such that the limits on the cross section are not strongly dependent on the mass.

It is worth noting that the systematic uncertainties affect smaller heavy neutrino masses more than larger ones. In particular, this effect is relevant when $M_N < $ 400 (600) GeV at the HL-LHC (FCC-hh). For 200 GeV mass, the limits can be weakened greatly by adding a 10\% systematic uncertainty on the background. Therefore, controlling the systematic uncertainty at the future pp colliders will be very important to enhance the discovery power for sterile neutrinos with small masses. 

The results presented here can also be representative for final states with the $\tau$ flavor.
In this case, additional backgrounds have to be included, and the difficulty of reconstructing the tau lepton has to be taken into account.
Consequently, we expect the 
sensitivities of the LNC-LFV $\tau^\pm \mu^\mp jj$ and $\tau^\pm e^\mp jj$ final states
to be weaker.
However, also the present constraints on the combinations 
$|V_{e N} V_{\tau N}|$ and $|V_{\mu N} V_{\tau N}|$
are much weaker compared to those from MEG on
$|V_{e N} V_{\mu N}|$.
The $\tau^\pm \mu^\mp jj$ and $\tau^\pm e^\mp jj$ channels could therefore mean great discovery potential, but require a dedicated analysis which 
is
left for future studies.

\begin{acknowledgments}
We thank the members of the FCC study collaboration for many useful discussions.
K.W. also wants to thank Christophe Grojean and Cai-dian L$\rm \ddot{u}$ for helpful discussions and their support.
This work has been supported by the Swiss National Science Foundation.
K.W.\ acknowledges support from the International Postdoctoral Exchange Fellowship Program (No.90 Document of OCPC, 2015).
O.F.\ acknowledges support from the ``Fund for promoting young academic talent'' from the University of Basel under the internal reference number DPA2354 and has received funding from the European Unions Horizon 2020 research and innovation program under the Marie Sklodowska-Curie grant agreement No 674896 (Elusives).

\end{acknowledgments}

\appendix

\section{Multivariate and Statistical Analysis}
\label{app:mva}
In this section we describe our setup of the Multivariate analysis (MVA), which is a statistical analysis of large data sets based on machine learning techniques to discriminate between two sets of data.
Here we use the Tool for MultiVariate Analysis (TMVA)~\cite{TMVA2007}, employing the Boosted Decision Tree (BDT).

We perform a frequentist test which uses the profile Likelihood ratio as test statistics. 
In addition to the parameters of interest for the limit calculation such as, the total cross section of the process and the integrated luminosity, we include nuisance parameters for background of $10\%$
to account for the unknown systematics at future colliders, assuming a logarithmic-normal distribution.

We construct an upper expected limit for the signal with upper/lower one and two sigma error bands using Higgs Analysis-Combined Limit tool~\cite{combine}. 
The limits can be set via the level of agreement between the data collected and a given hypothesis by computing the probability of finding the observed data incompatible with the prediction for a given hypothesis, this probability is referred to as the $p$-value.

The expected value of finding the number of events in the $i$th bin of the BDT distribution is given by

\begin{equation}
E[n_i] = \mu S_i + B_i\,,
\end{equation}
where the parameter $\mu$ is called the signal strength.
When a hypothesis with $\mu=0$ is rejected a discovery can be established, while rejecting the hypothesis with $\mu=1$ defines our limit for the calculation. 
The likelihood function is constructed as Poisson probabilities for all bins as:
\begin{equation}
{\cal{L}}(\mu,\theta) = \sum_{i=1}^n\frac{(\mu S_i + B_i)^{n_i}}{n_i!}e^{-(\mu S_i + B_i)}.
\end{equation}
The profile likelihood ratio can be constructed by the Maximum-Likelihood Estimate (MLE) as:
\begin{equation}
\lambda(\mu) = \frac{{\cal{L}}(\mu,\hat{\theta} )}{{\cal{L}}(\hat{\mu},\hat{\theta})}
\end{equation}

with ${\hat{\theta}}$ and ${\hat{\mu}}$ are the estimated parameters for $\theta$ and $\mu$ that maximize the likelihood function, i.e., for a given $\mu$ and pseudo data at $\hat{\theta}$, the combined $\hat{\mu}$ with $\hat{\theta}$ define the point for which the likelihood reaches its global maximum. The fact that the profile likelihood ratio depends on systematical errors broaden the estimate of the maximum likelihood, thus large systematical errors lead to weaker limits. For the statistical test one can construct the profile log likelihood as:
\begin{equation}
q(\mu) = -2\ln\lambda(\mu)
\end{equation}
To measure the level of incompatibility we compute $p-$value as:
\begin{equation}
p = \int_{q(\mu)}^\infty F[q(\mu)|\mu] dq(\mu),
\end{equation}
with $F[q(\mu)|\mu]$ being the probability distribution function that measures the incompatibility between data and our hypothesis, while higher values of $q(\mu)$ correspond to high disagreement between data and hypothesis. 
The signal is excluded at $(1-\alpha)$ confidence level if
\begin{equation}
CL_s = \frac{P\left( q(\mu) | \mu S + B \right)}{P\left( q(\mu) | B \right)} < \alpha ,
\end{equation}
where the upper limit on $\mu$ is the largest value for $\mu$ with $P < \alpha$, i.e., if $\alpha = 0.05$ then the signal is excluded with $95\%$ confidence level. Thus one can simply get the upper exclusion limit at different confidence levels by
\begin{equation}
\mu_{up} = \hat{\mu} + \sigma\Phi^{-1}(1-\alpha)
\end{equation}

with $\hat{\mu}$ being the estimated expected median and $\Phi^{-1}$ being a cumulative distribution function. We use the following confidence levels: 
$(1-\alpha) = 0.6827$ corresponds to the $1-\sigma$ confidence level; $(1-\alpha) = 0.9545$ corresponds to the $2-\sigma$ confidence level; $(1-\alpha) = 0.997$ corresponds to the $3-\sigma$ confidence level and $ (1-\alpha) = 0.9999$ corresponds to the $5-\sigma$ confidence level. Finally, the error bands can be obtained by
\begin{equation}
{\rm Band}_{(1-\alpha)}= \hat{\mu} \pm \frac{\sigma\Phi^{-1}\left( 1-\alpha \right)}{N}\;.
\end{equation}

In fact if we restrict the number of events for the signal and the background to be large and ignore the correlation effect between bins, one can calculate the limit from the following formula for the significance
\begin{equation}
\sigma_{\rm stat+syst} =
\Bigg[ 2 \bigg( (N_s + N_b) {\rm ln} \frac{(N_s+N_b)(N_b+\sigma_b^2)}{N_b^2+(N_s+N_b)\sigma_b^2} - \frac{N_b^2}{\sigma_b^2} {\rm ln}(1+ \frac{\sigma_b^2 N_s}{N_b(N_b+\sigma_b^2) } )\, \bigg)\, \Bigg]^{1/2}
\label{eqn:sgf2}
\end{equation}
with $N_s$, $N_b$ being the number of signal and background events, respectively, and $\sigma_b$ parametrising the systematic uncertainty.

\newpage
\section{Distributions of Input Observables}
\label{app:distributions}

\begin{figure}[h]
\centering
\includegraphics[width=4.5cm,height=3.5cm]{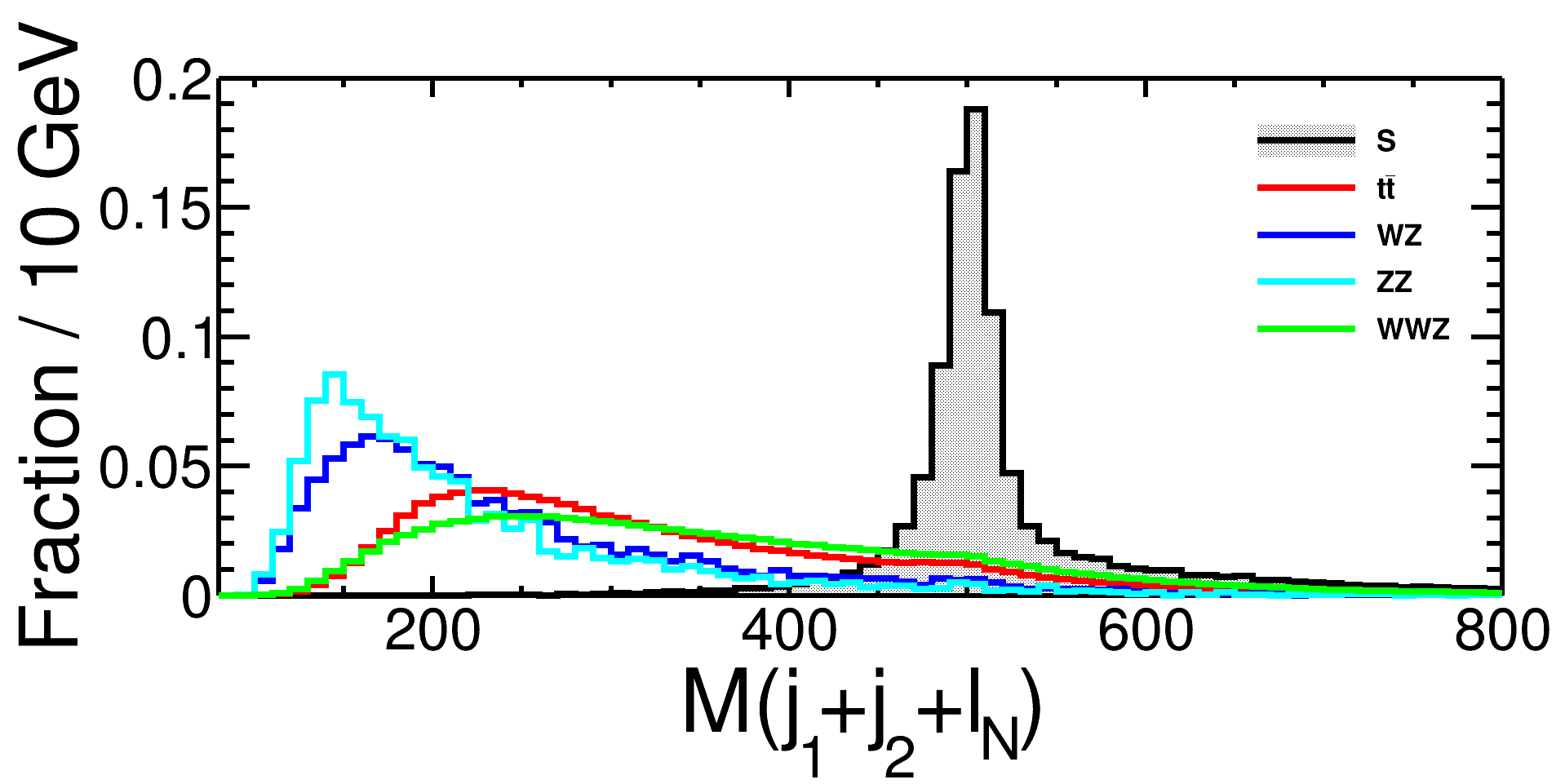}
\includegraphics[width=4.5cm,height=3.5cm]{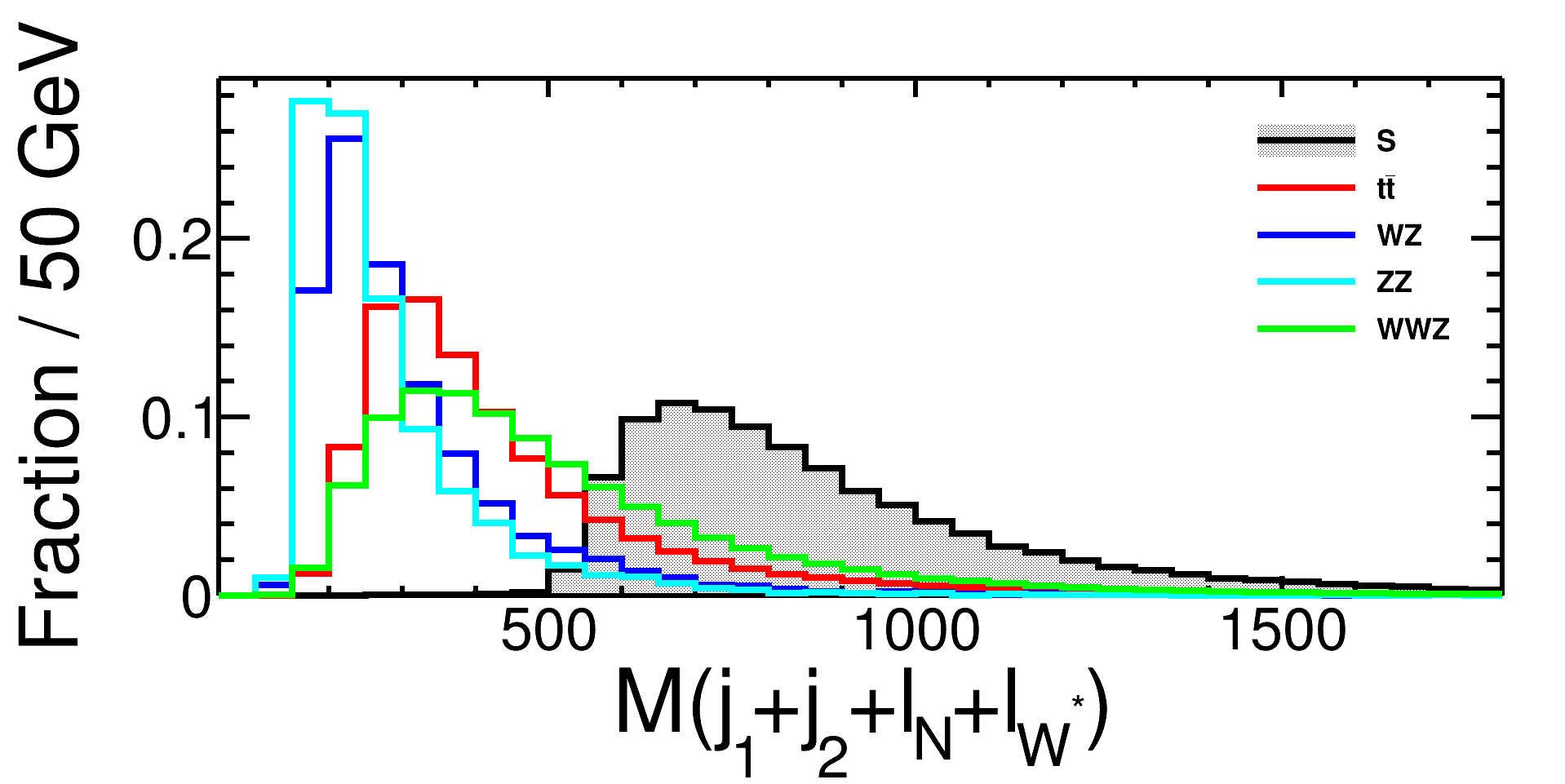}
\includegraphics[width=4.5cm,height=3.5cm]{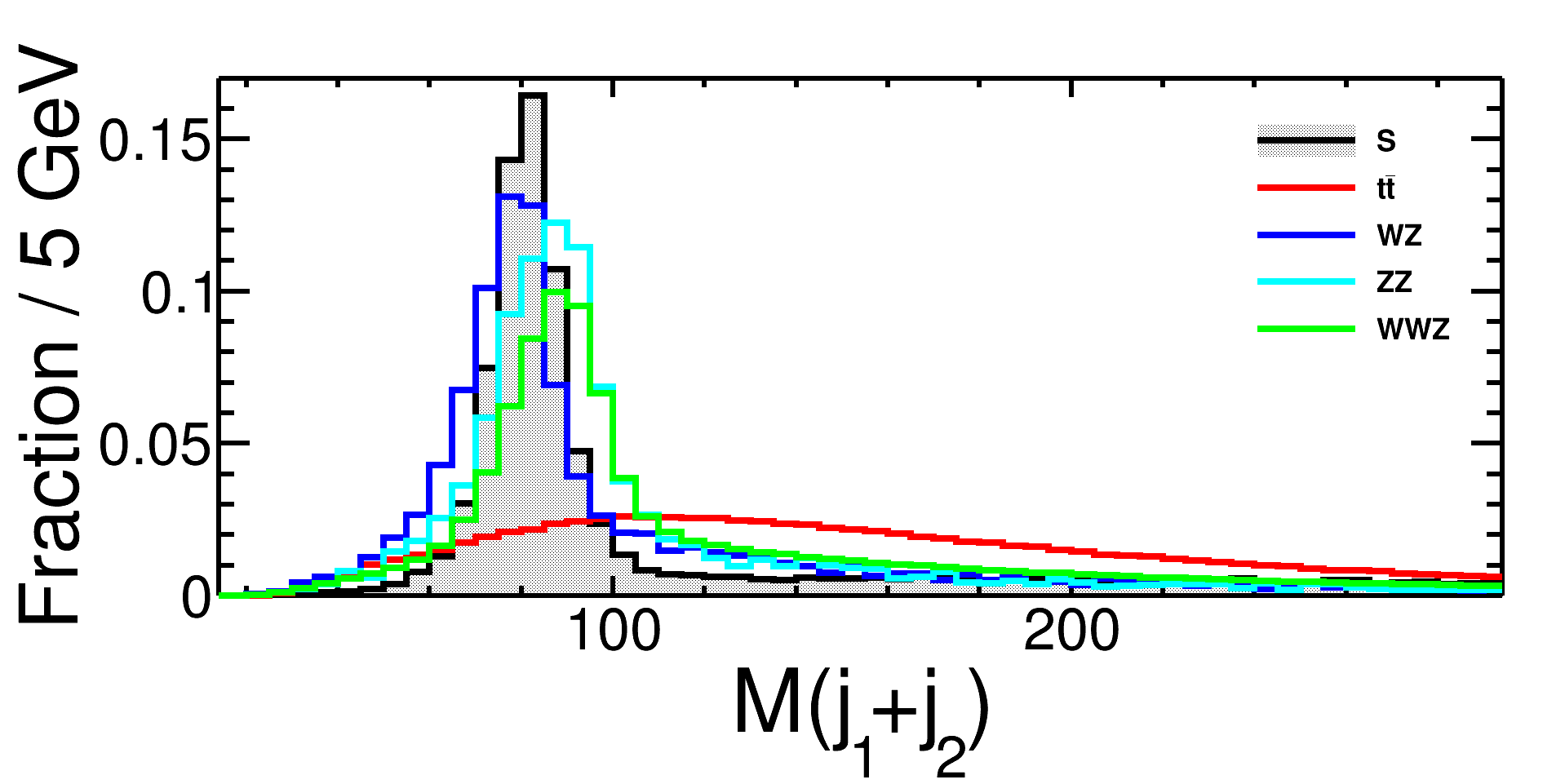}
\includegraphics[width=4.5cm,height=3.5cm]{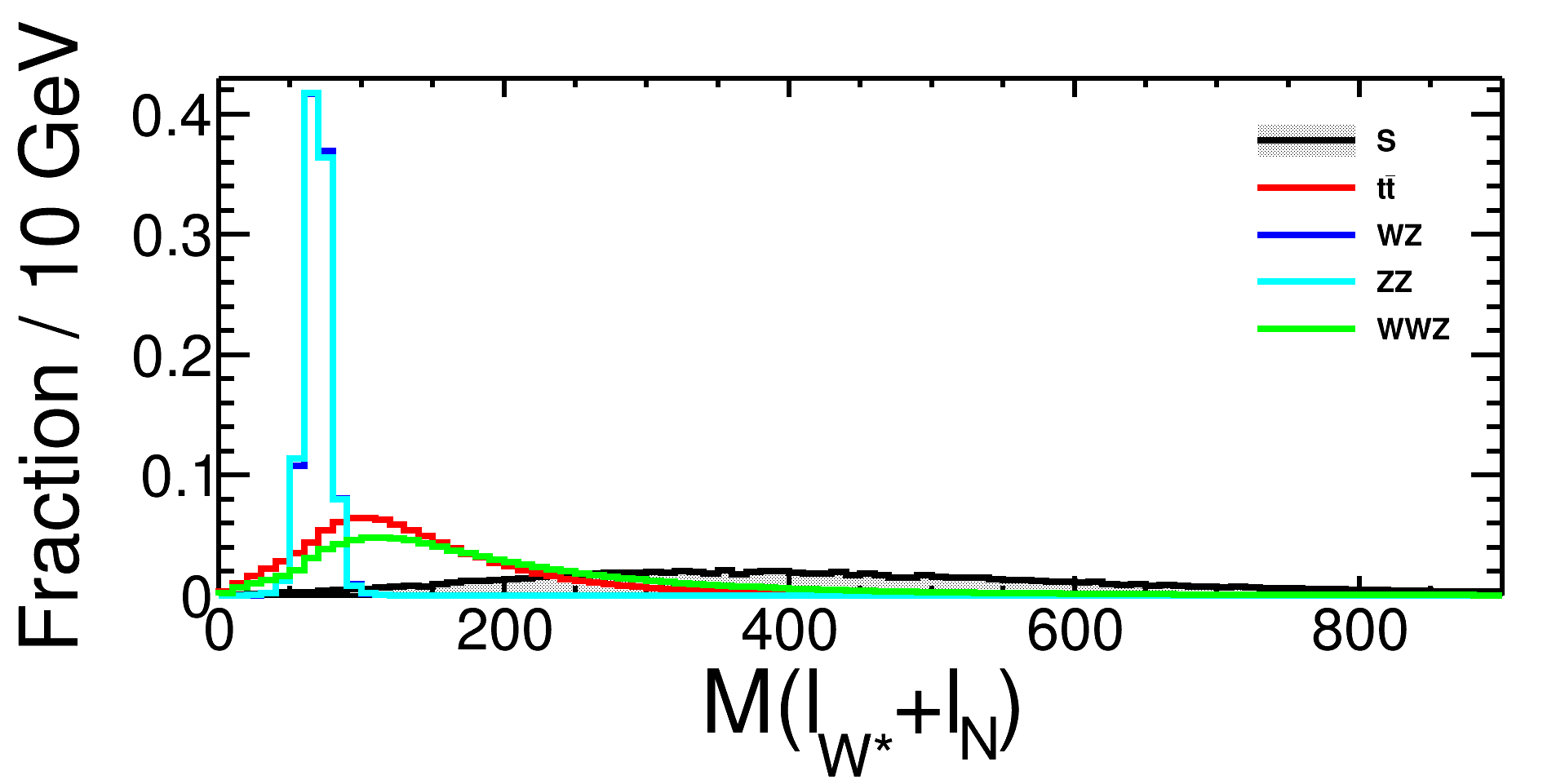}
\includegraphics[width=4.5cm,height=3.5cm]{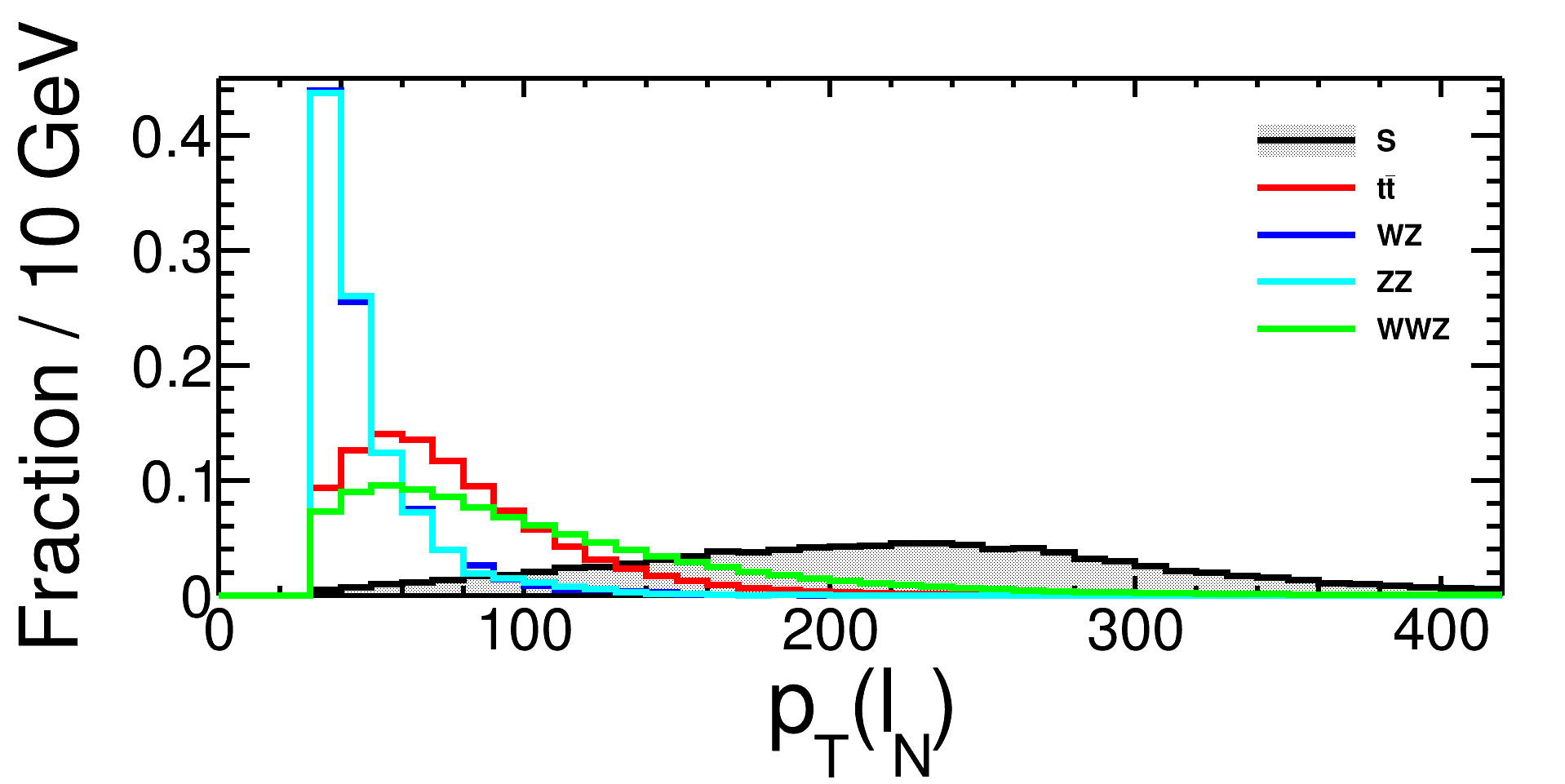}
\includegraphics[width=4.5cm,height=3.5cm]{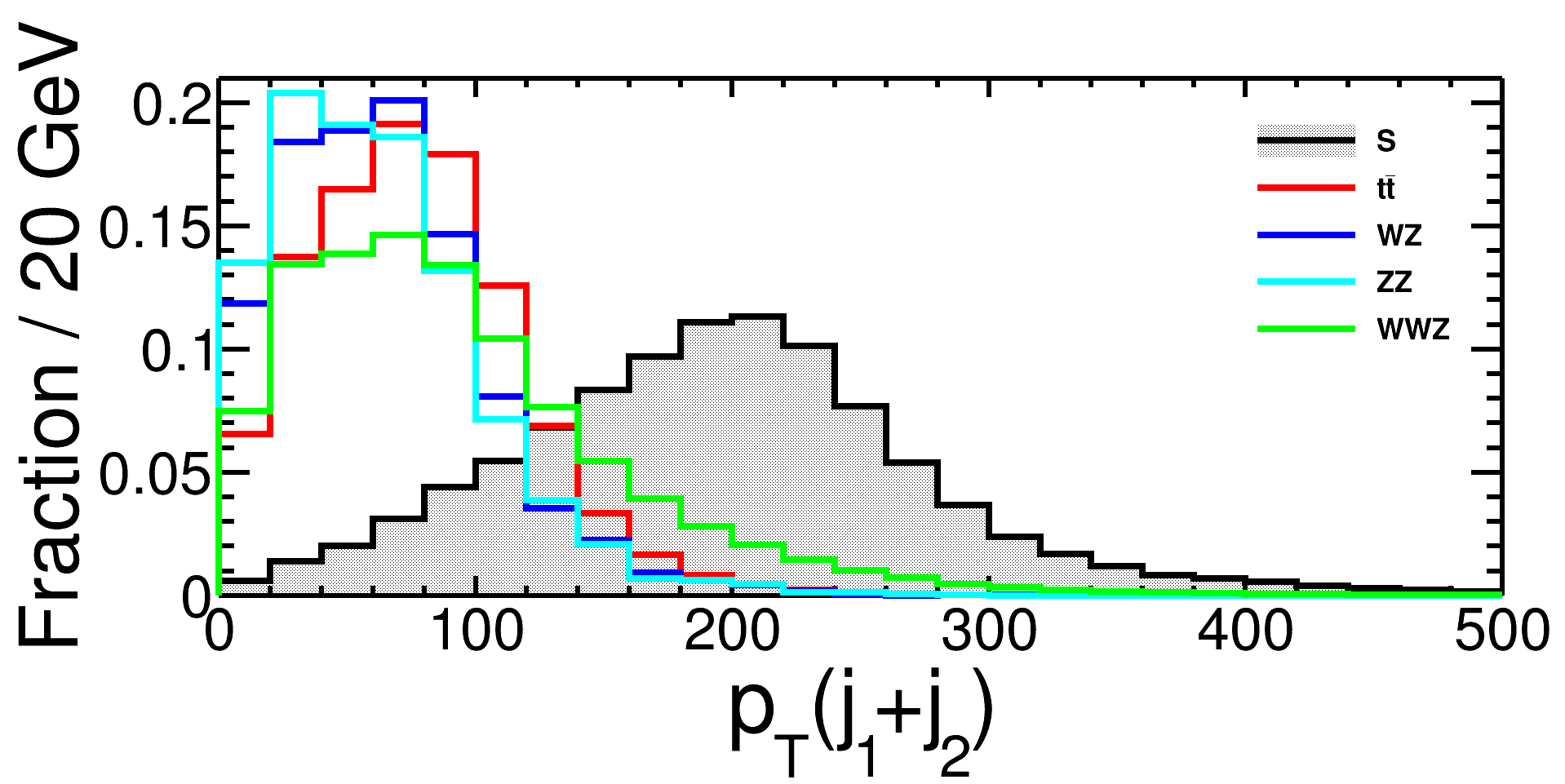}
\includegraphics[width=4.5cm,height=3.5cm]{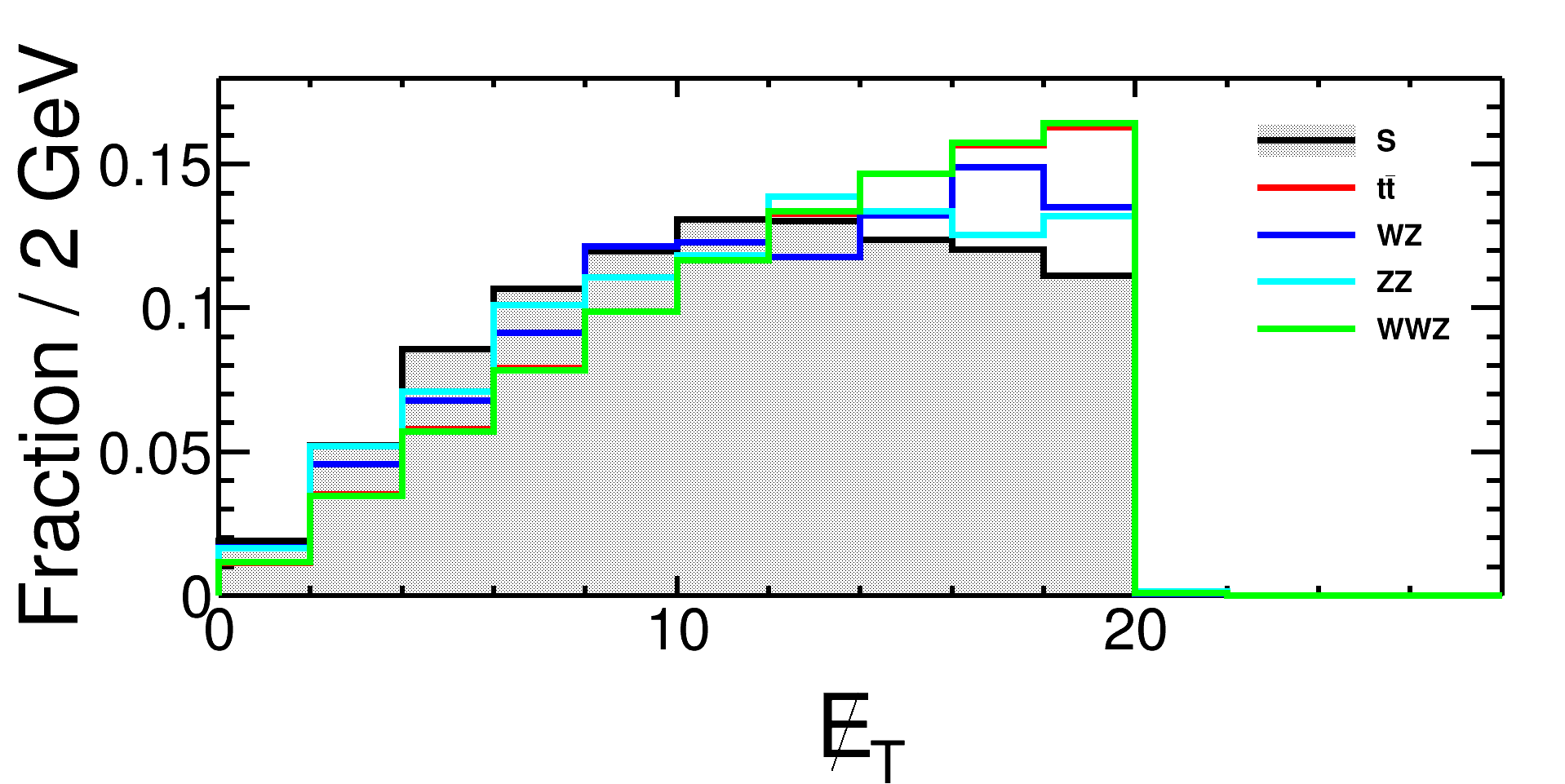}
\includegraphics[width=4.5cm,height=3.5cm]{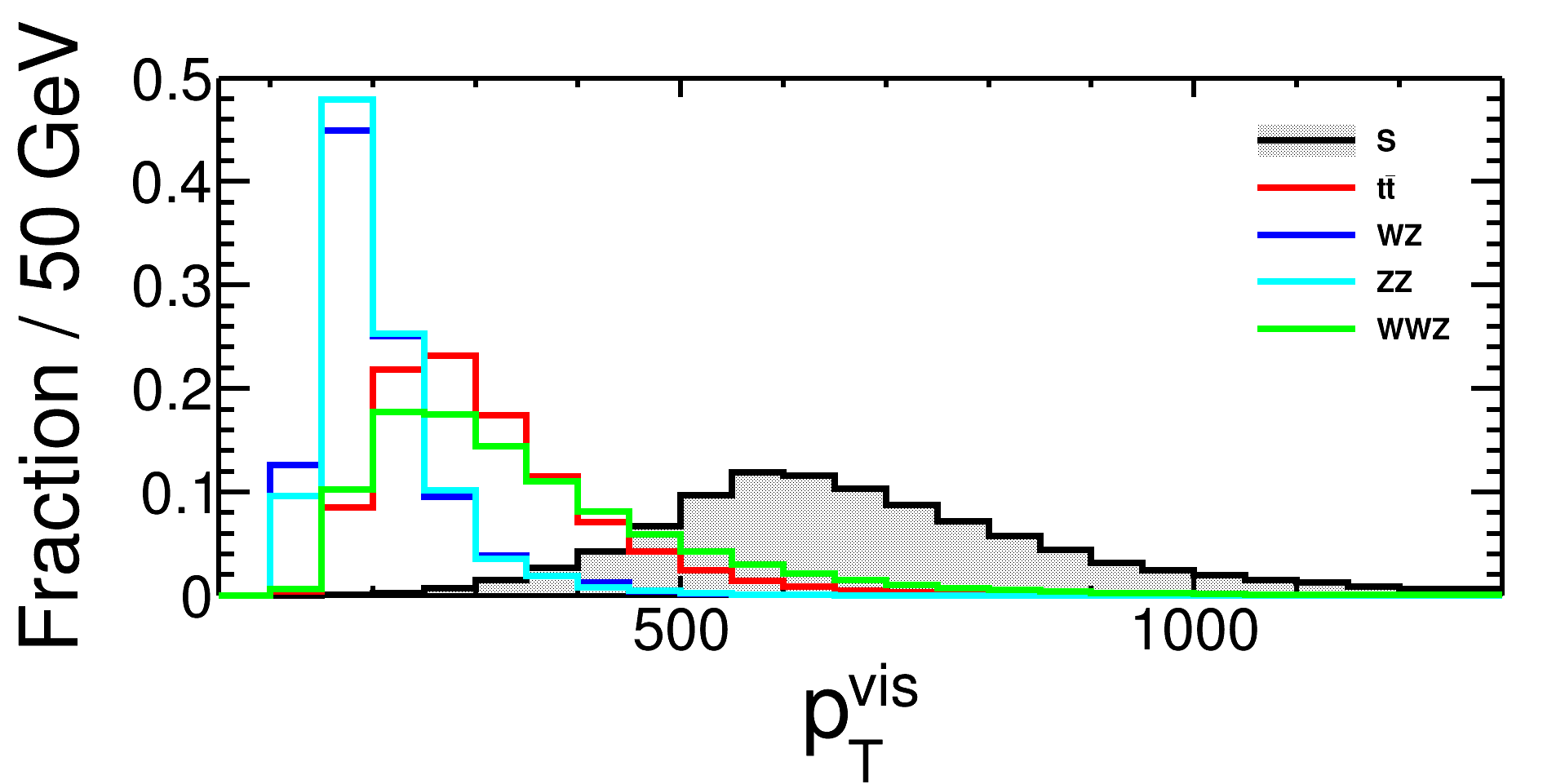}
\includegraphics[width=4.5cm,height=3.5cm]{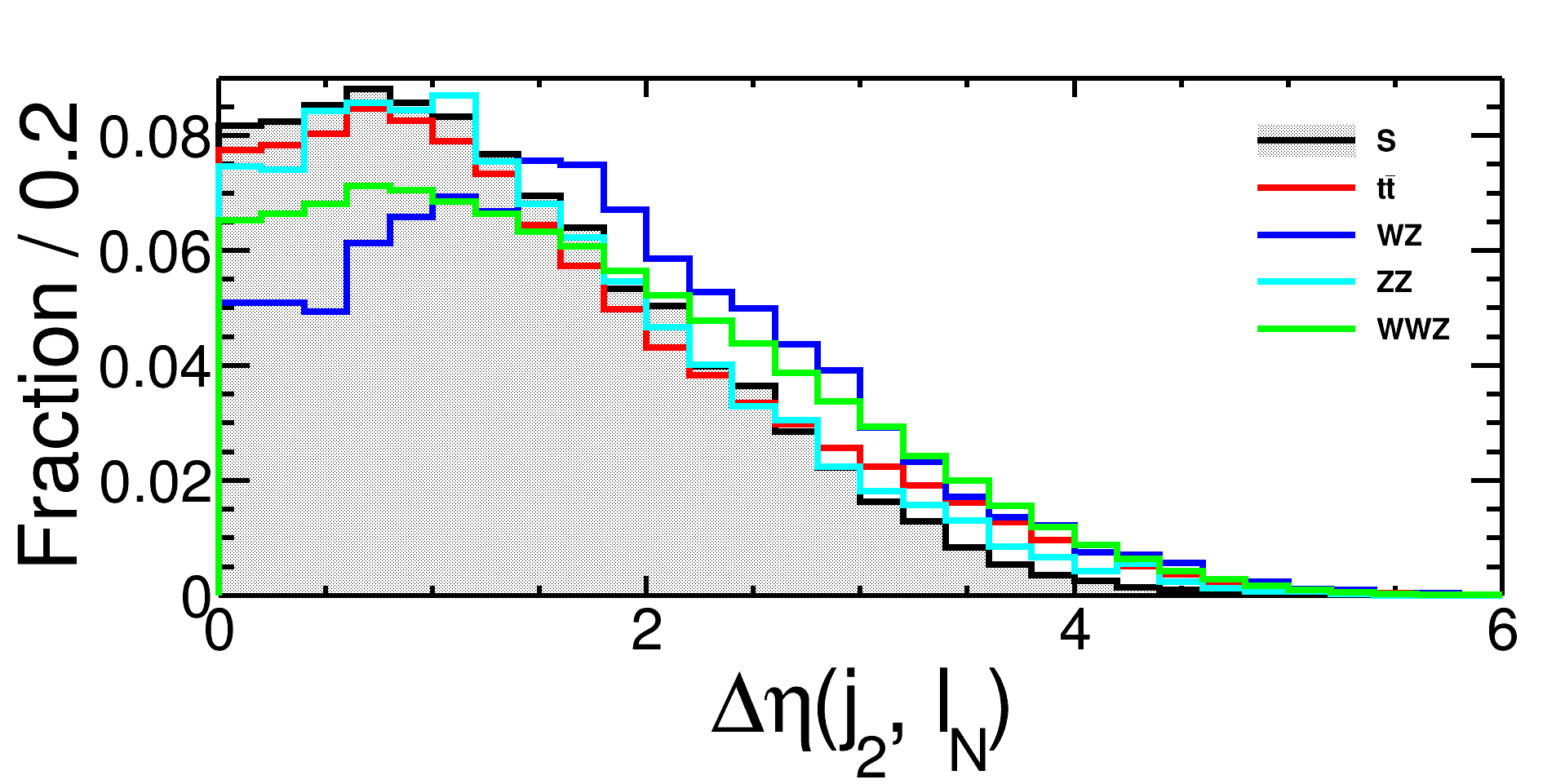}
\includegraphics[width=4.5cm,height=3.5cm]{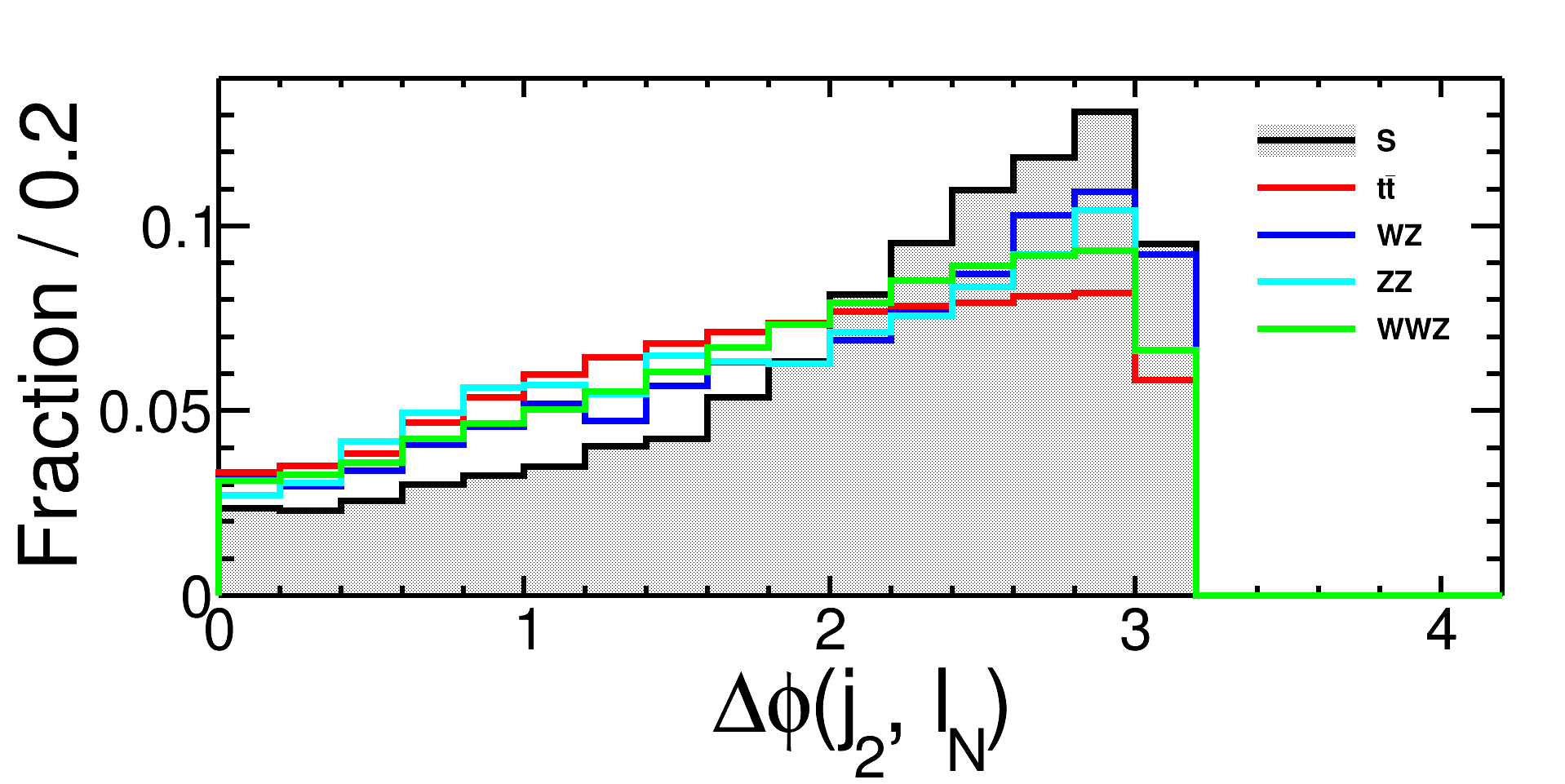}
\includegraphics[width=4.5cm,height=3.5cm]{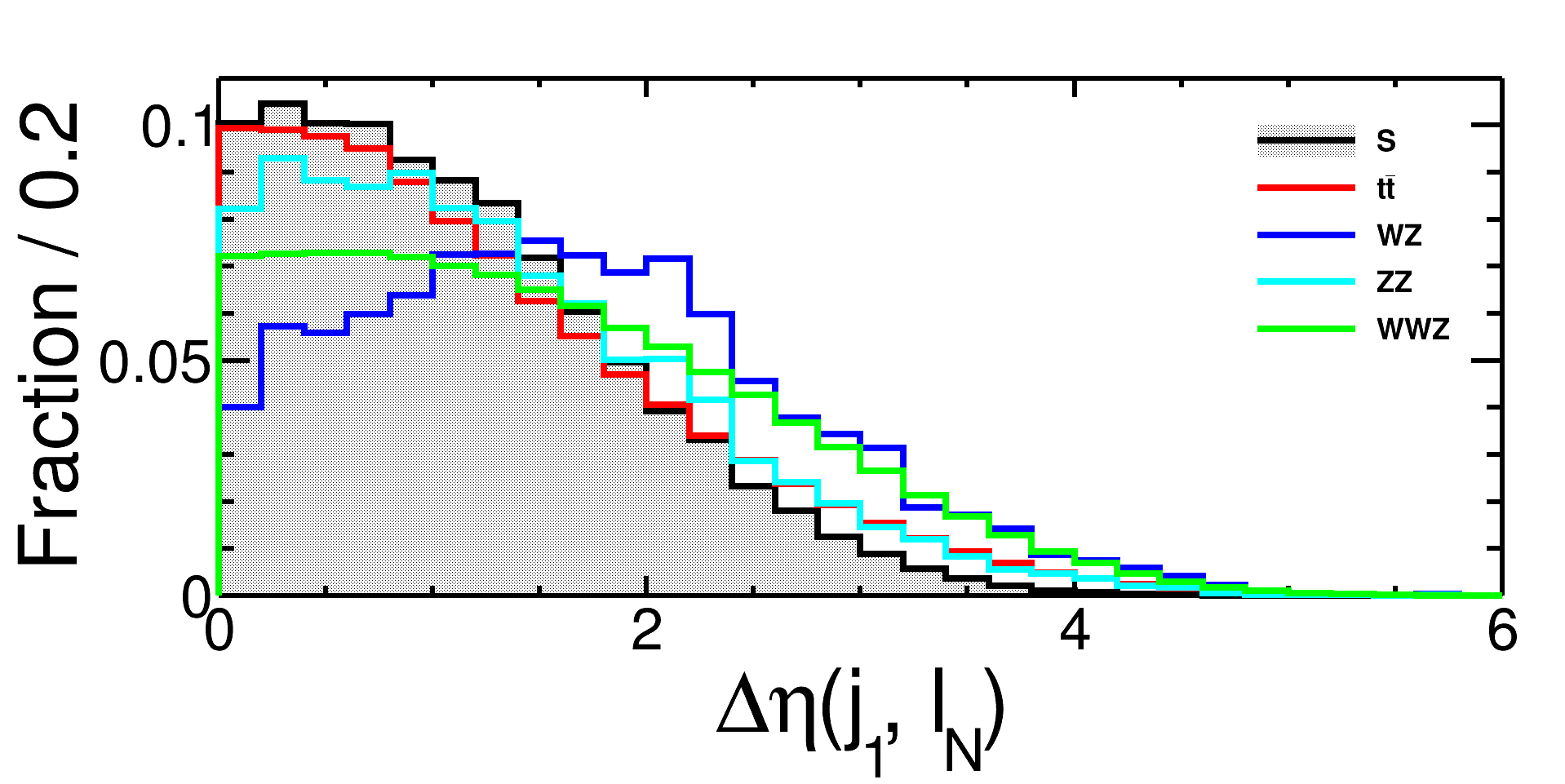}
\includegraphics[width=4.5cm,height=3.5cm]{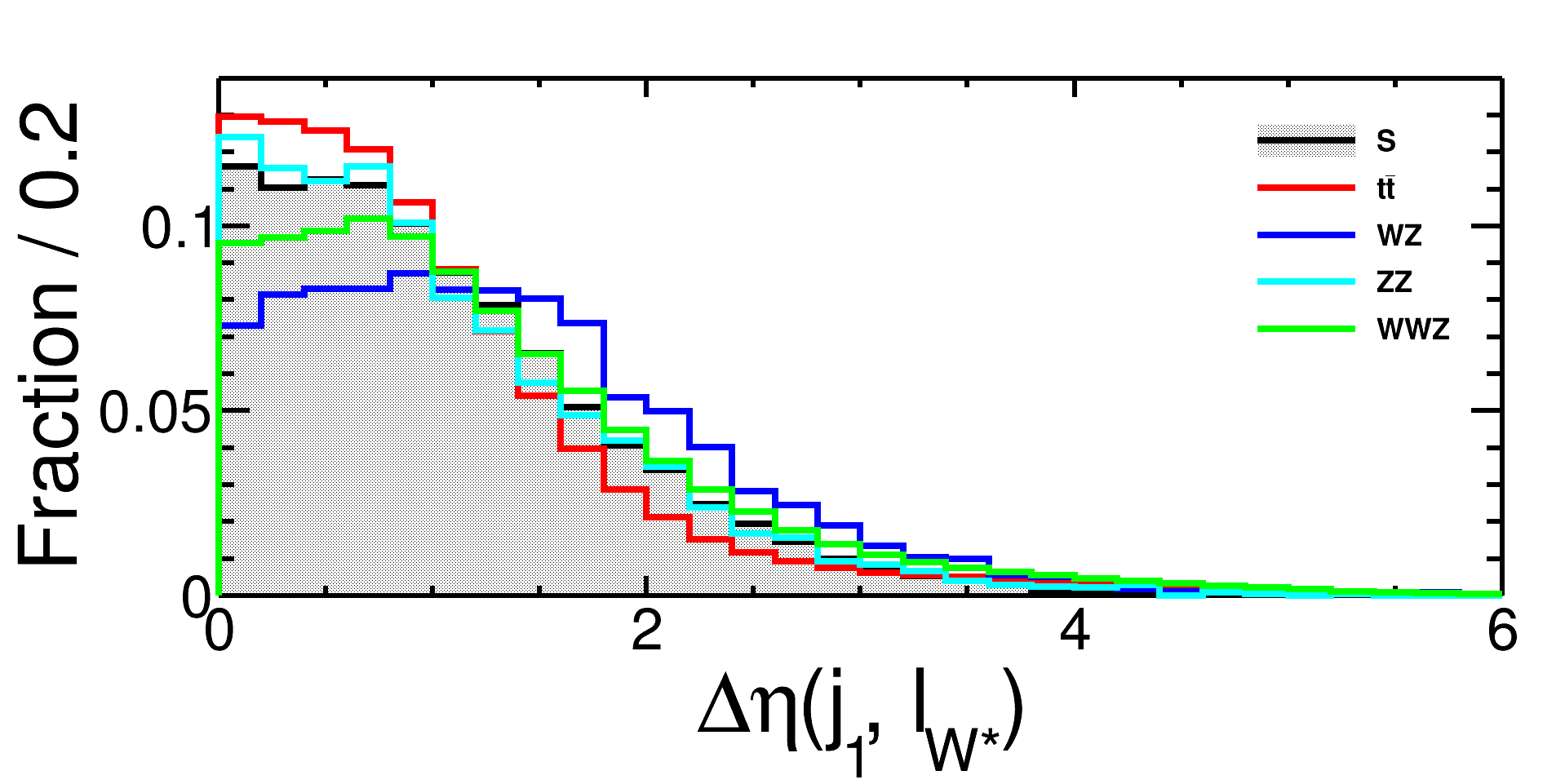}
\caption{Kinematic distributions of some selected observables for the signal
with $M_N$ = 500 GeV (S, black with filled area), and for SM background processes of $t\bar{t}$ (red), WZ (blue), ZZ (cyan), and WWZ (green) after applying the pre-selection cuts at the HL-LHC.}
\label{fig:HLLHC_distributions}
\end{figure}

\begin{figure}[h]
\centering
\includegraphics[width=4.5cm,height=3.5cm]{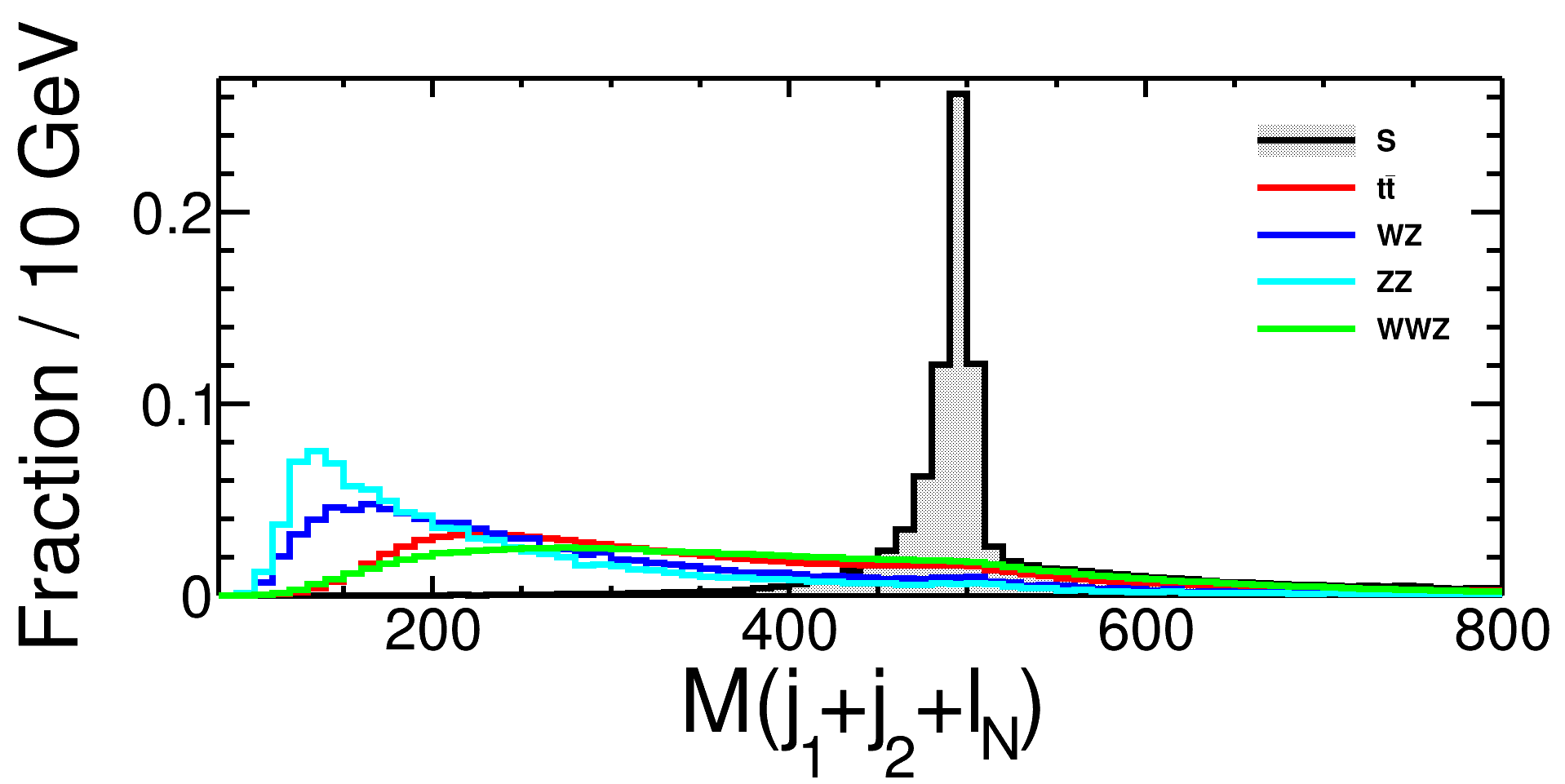}
\includegraphics[width=4.5cm,height=3.5cm]{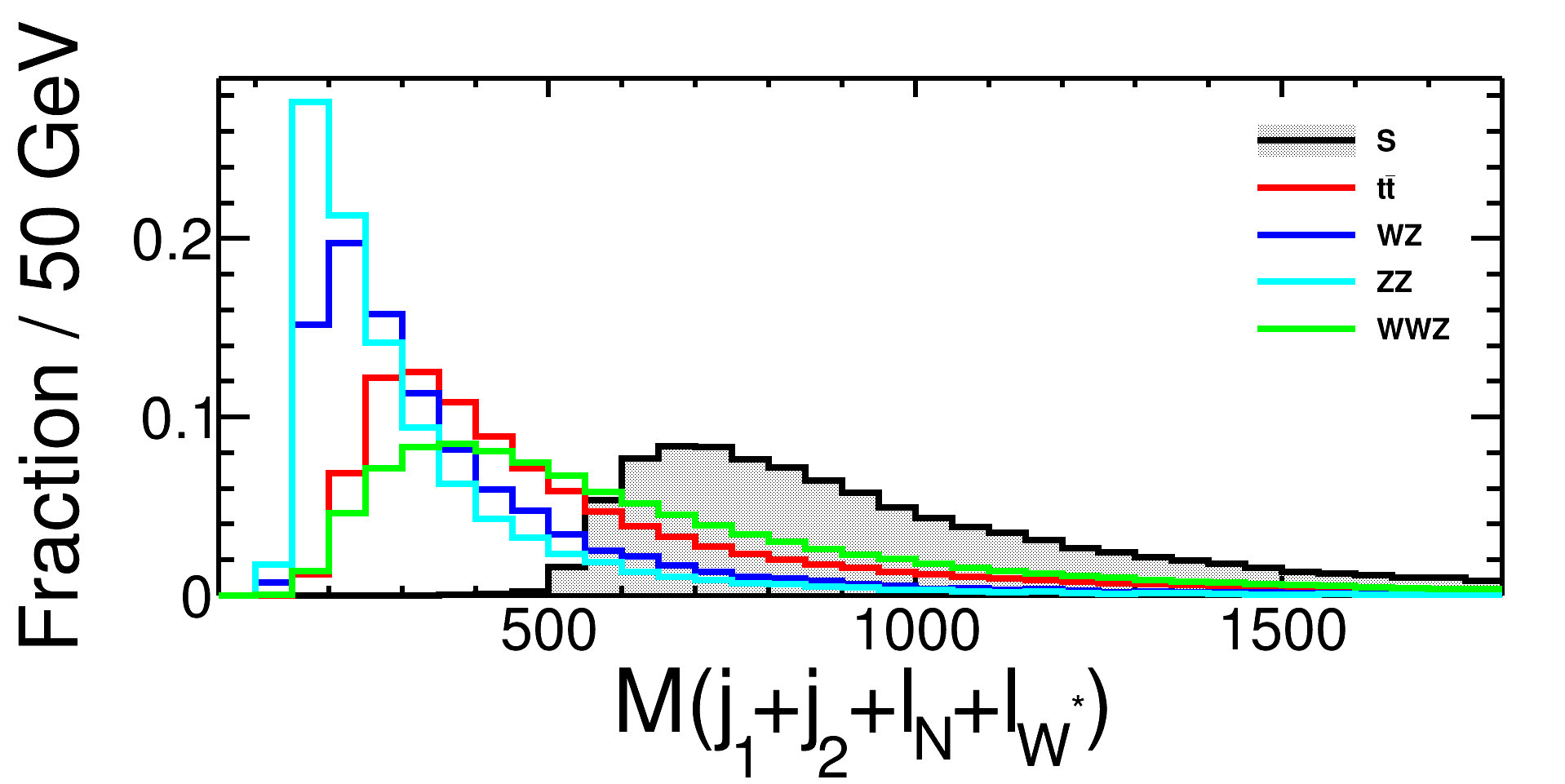}
\includegraphics[width=4.5cm,height=3.5cm]{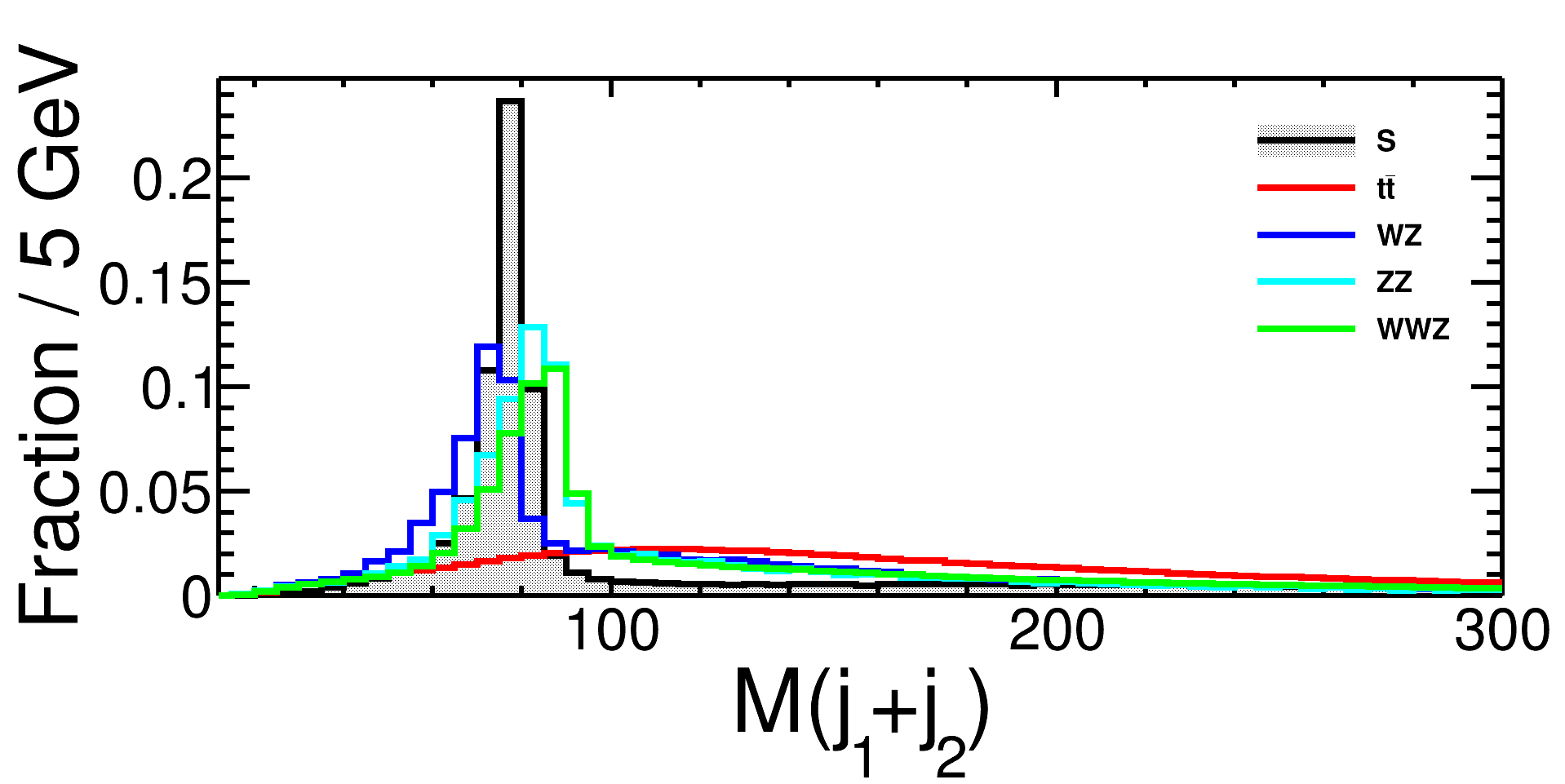}
\includegraphics[width=4.5cm,height=3.5cm]{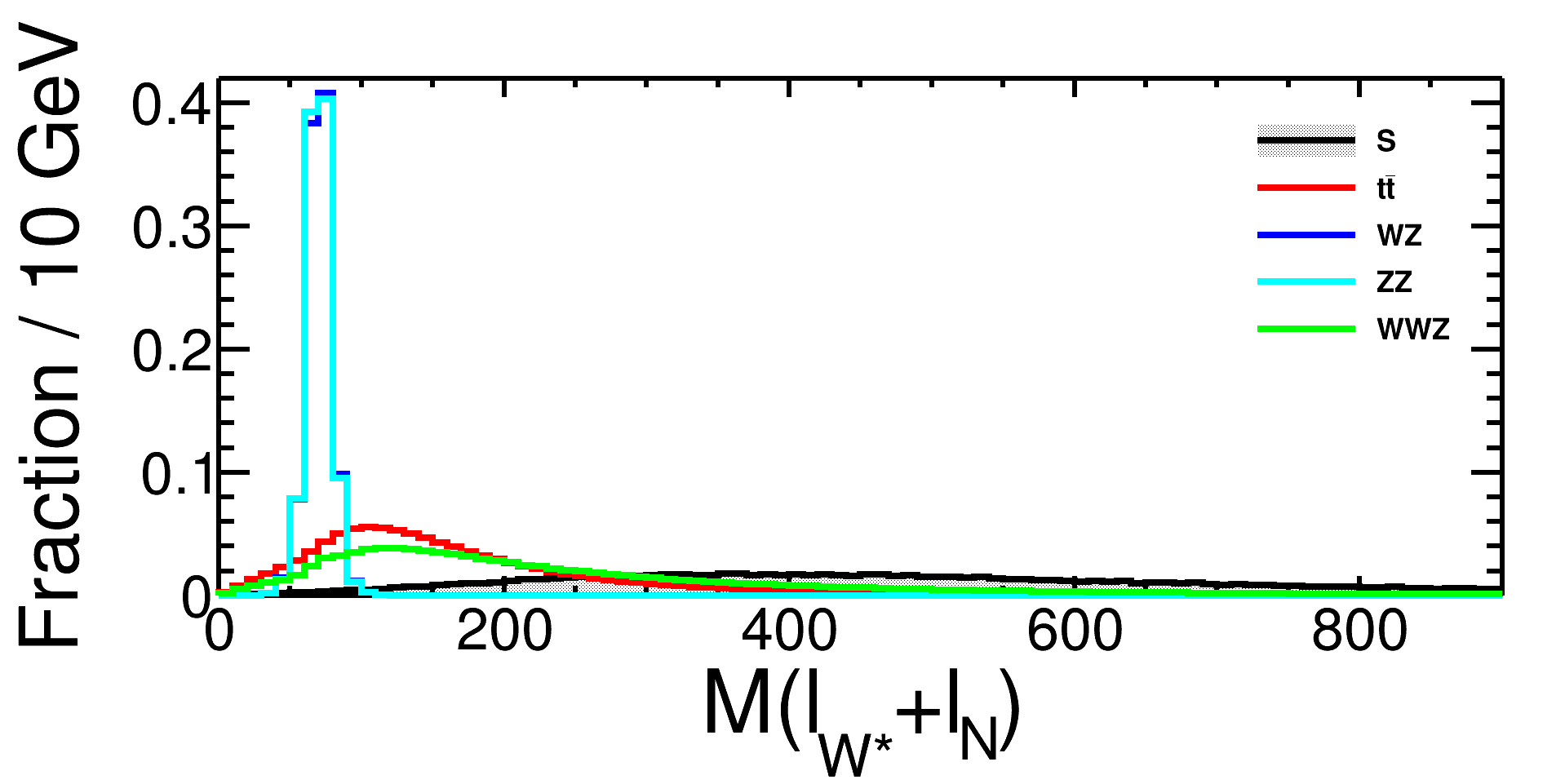}
\includegraphics[width=4.5cm,height=3.5cm]{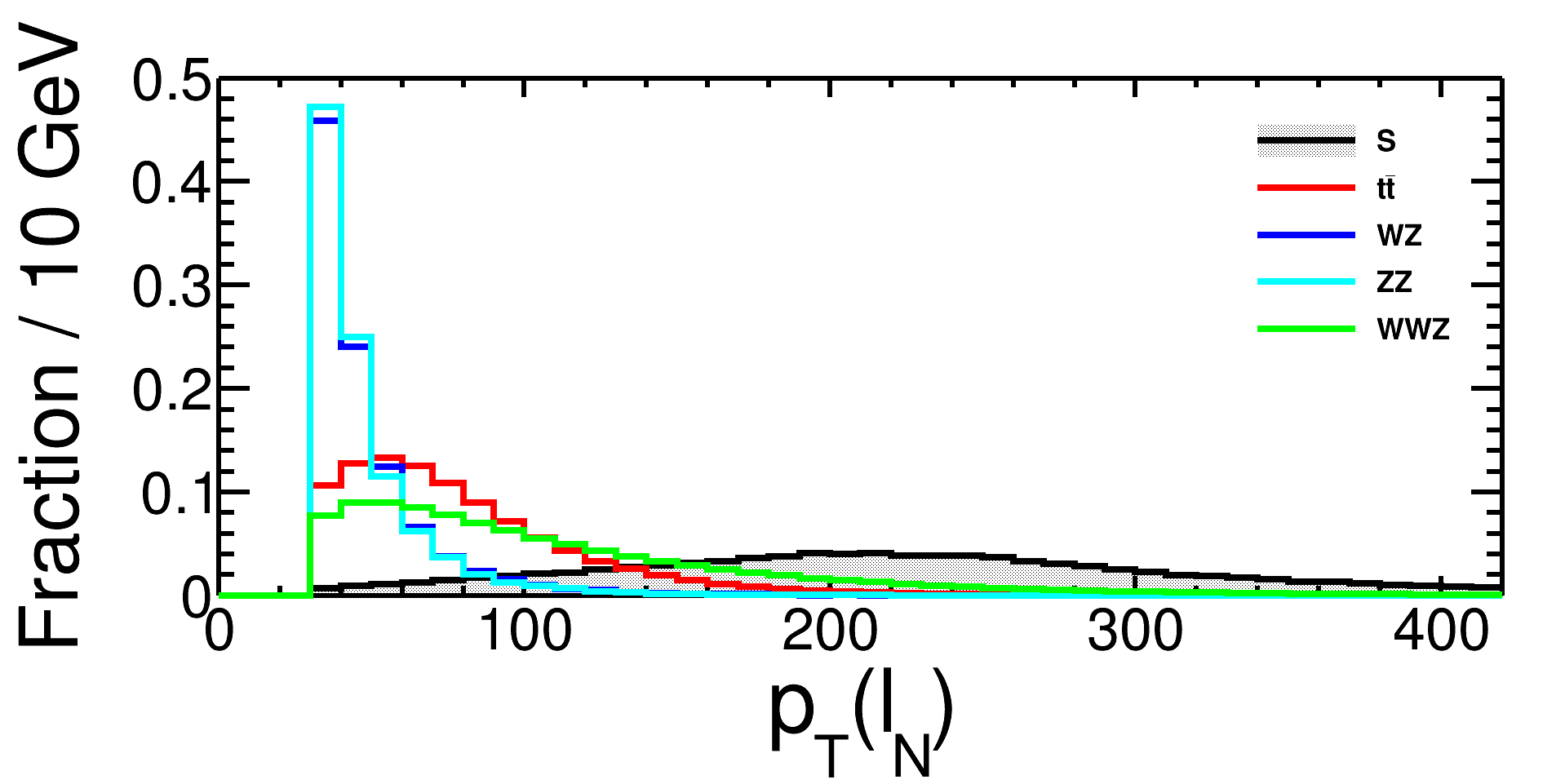}
\includegraphics[width=4.5cm,height=3.5cm]{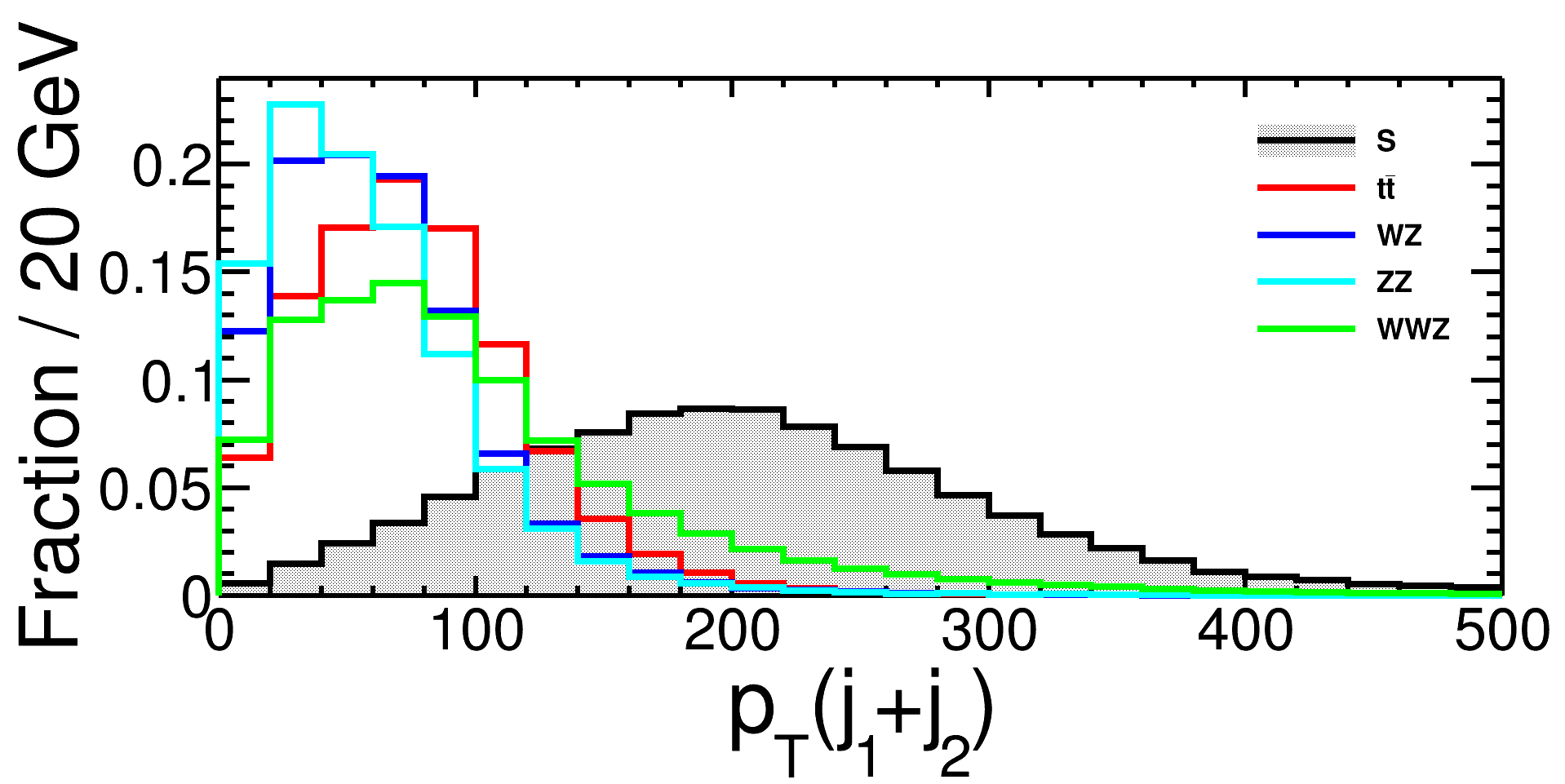}
\includegraphics[width=4.5cm,height=3.5cm]{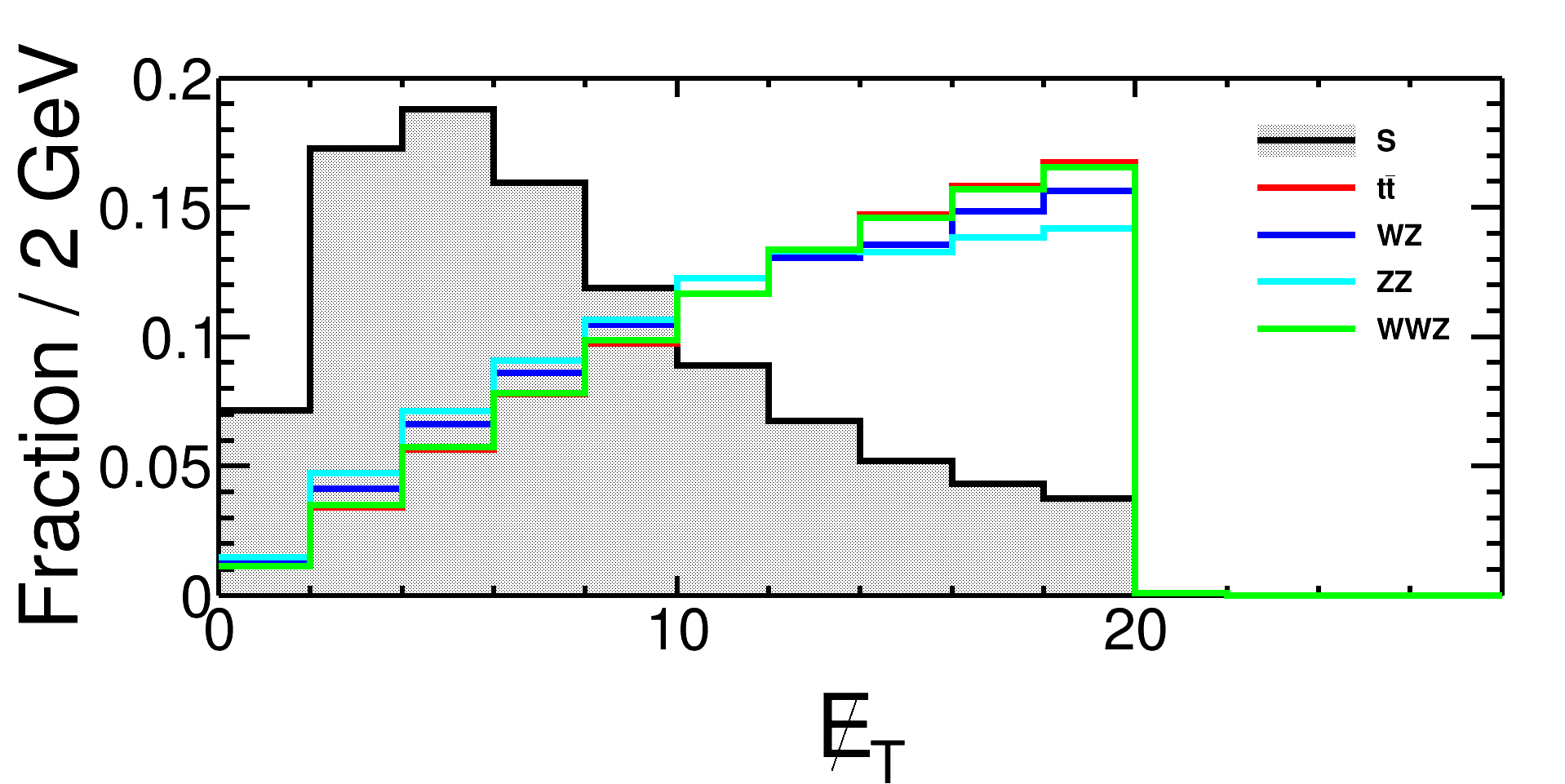}
\includegraphics[width=4.5cm,height=3.5cm]{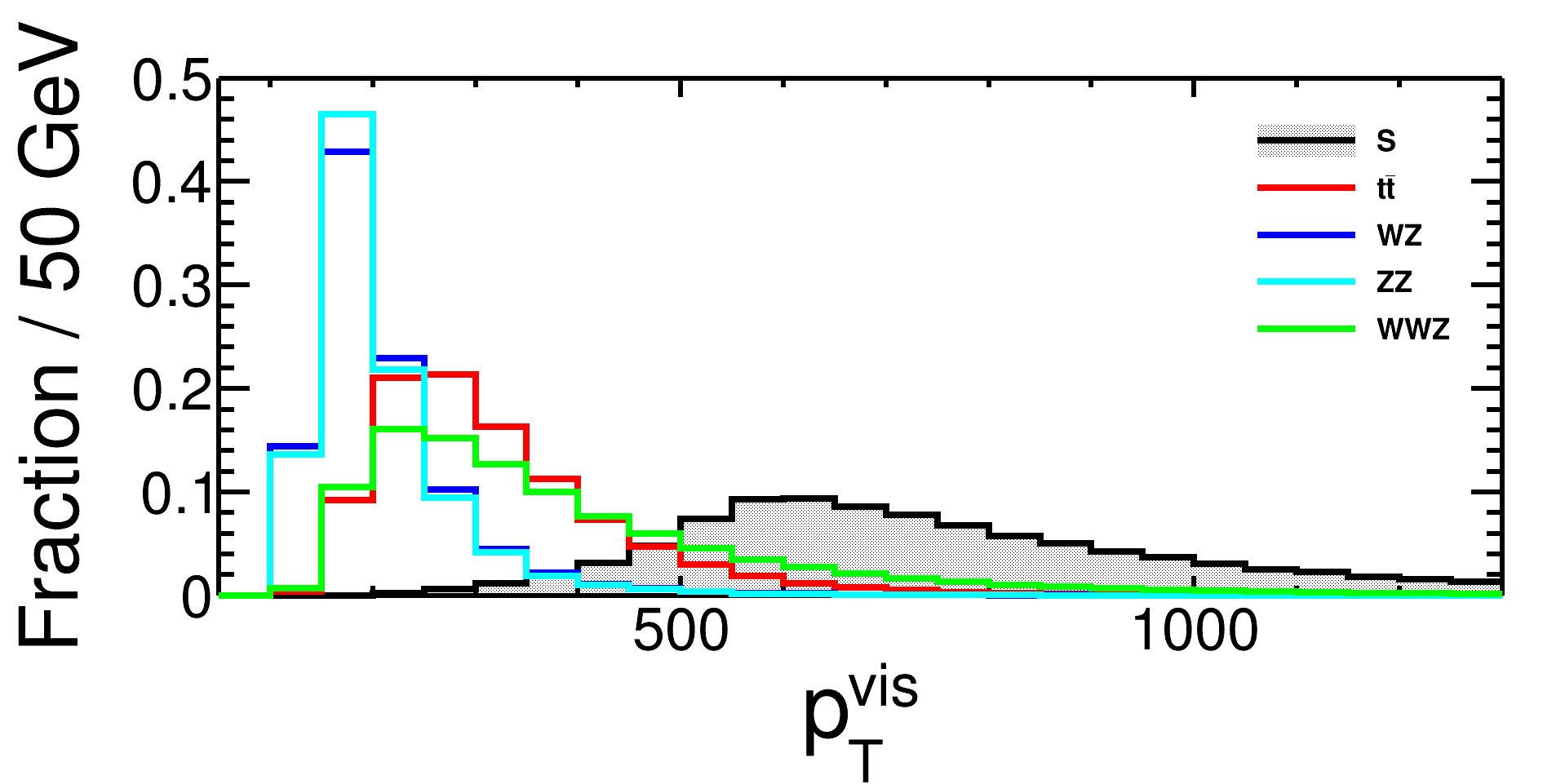}
\includegraphics[width=4.5cm,height=3.5cm]{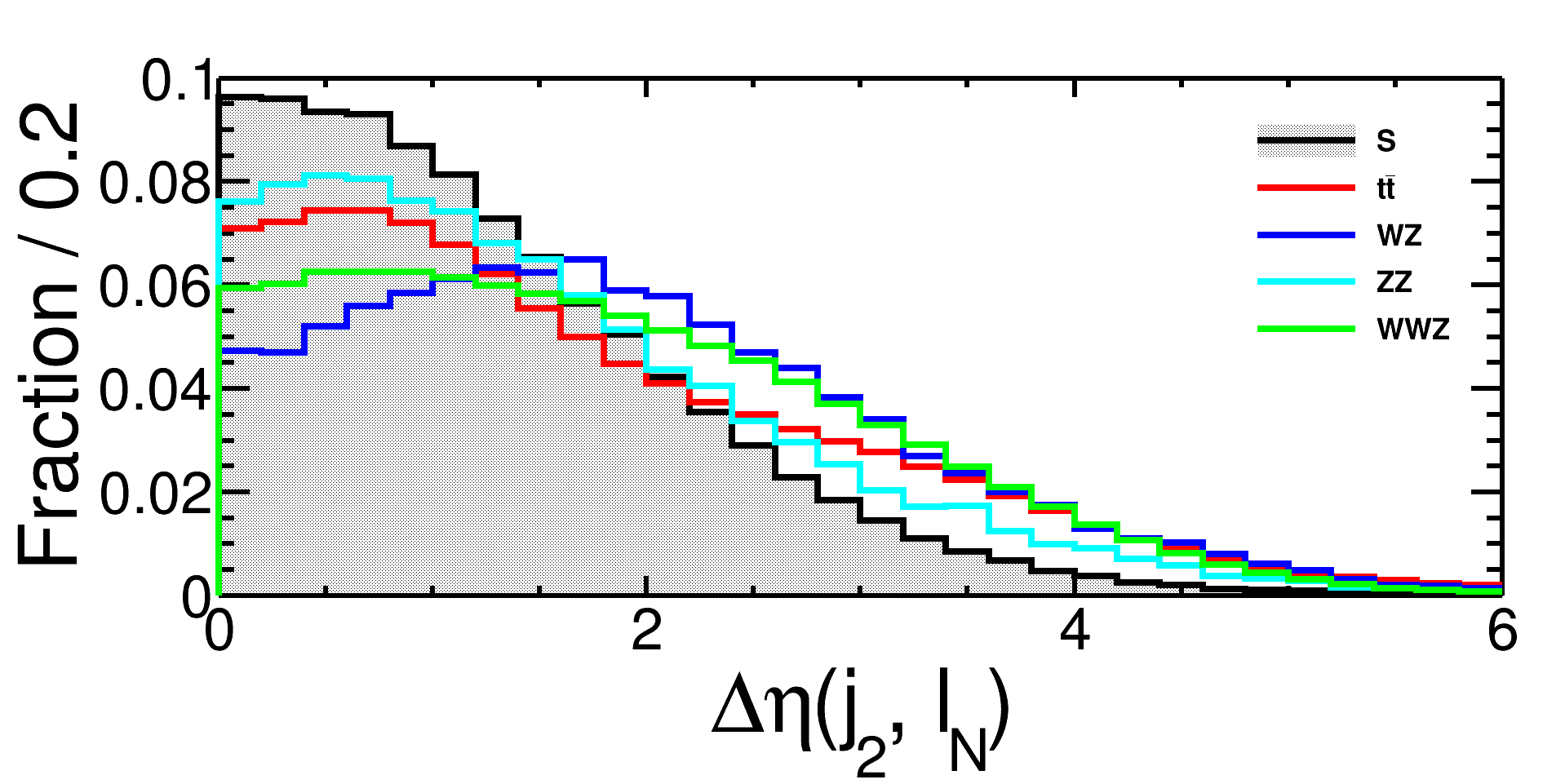}
\includegraphics[width=4.5cm,height=3.5cm]{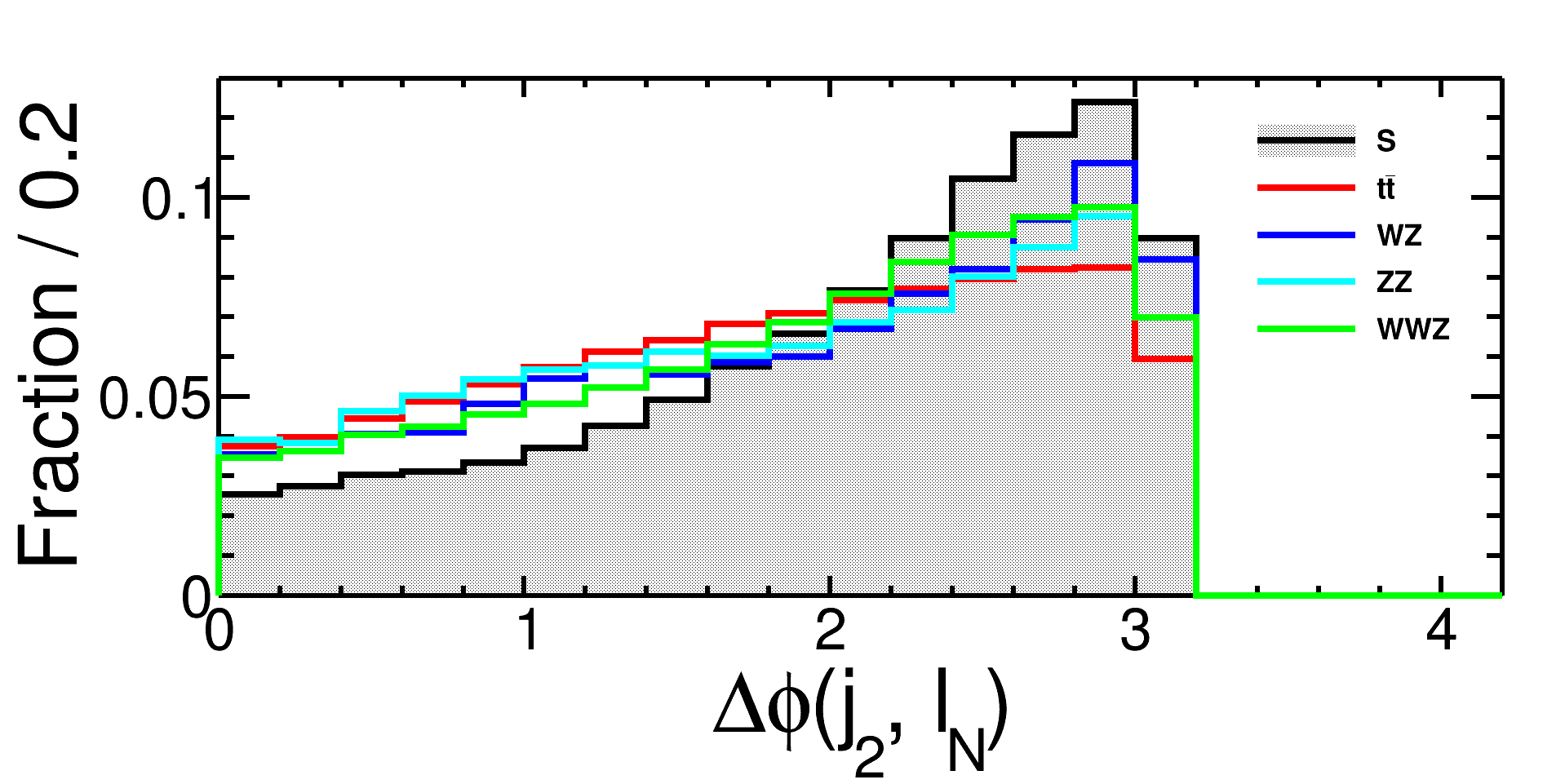}
\includegraphics[width=4.5cm,height=3.5cm]{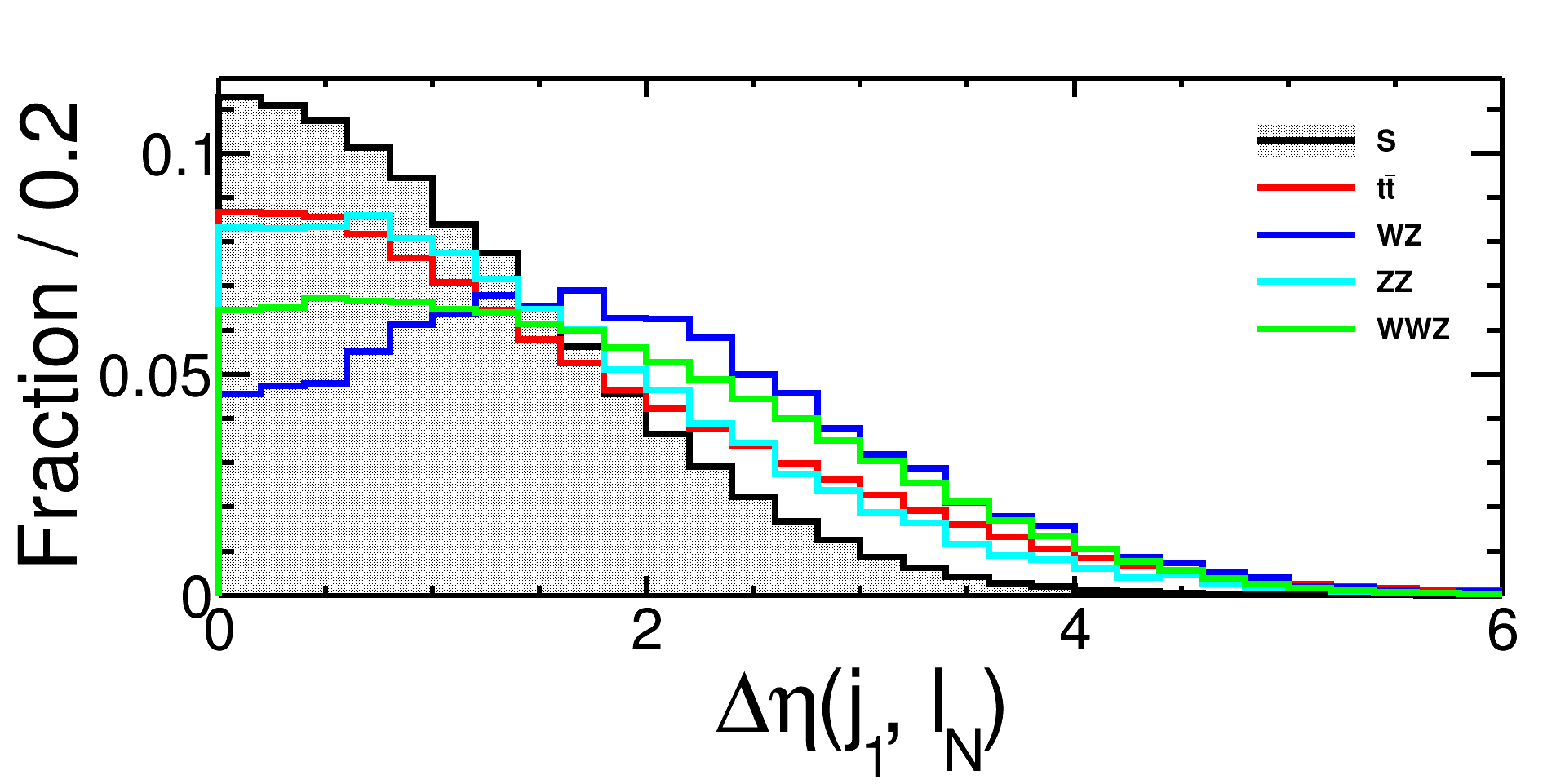}
\includegraphics[width=4.5cm,height=3.5cm]{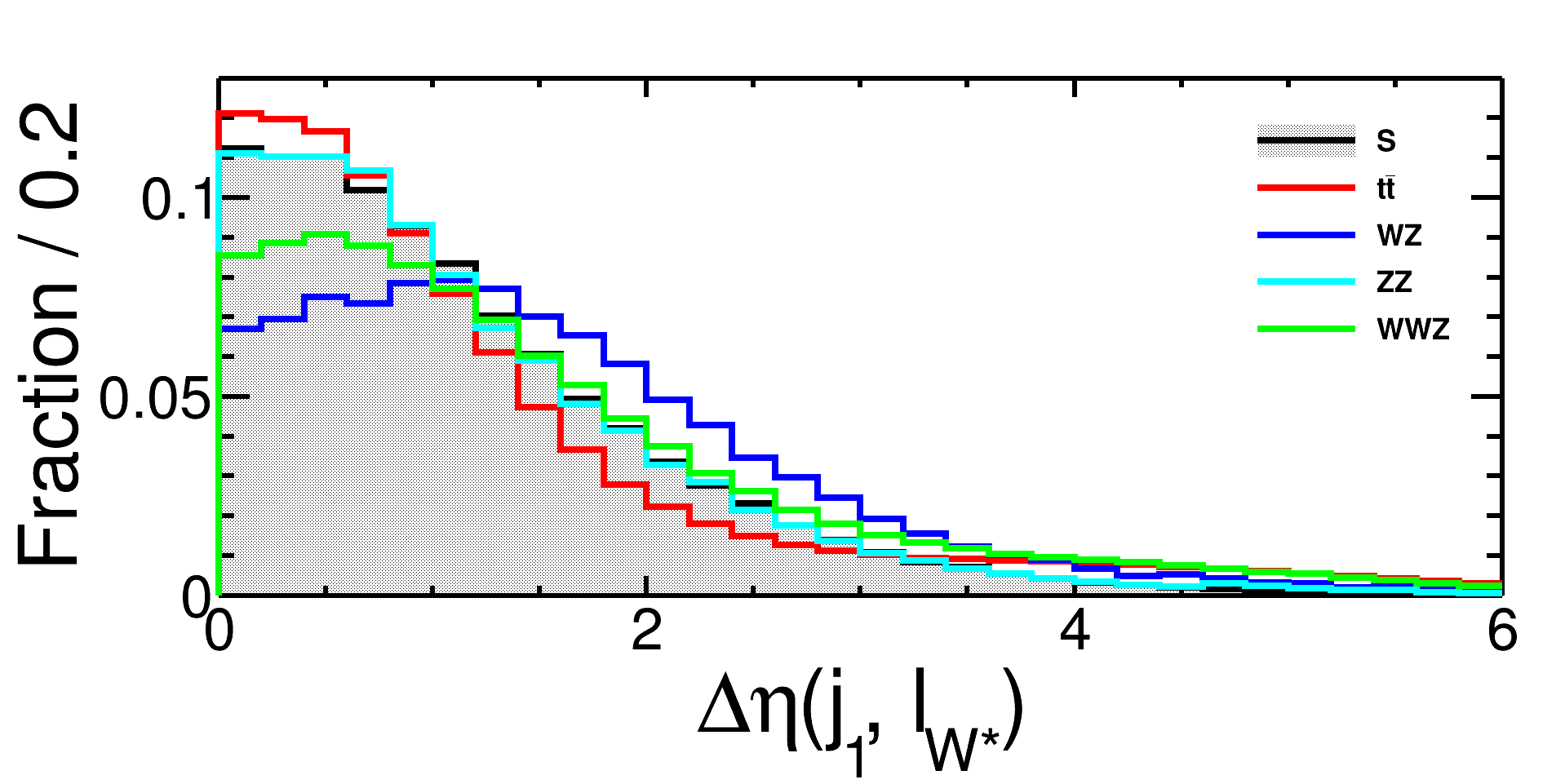}
\caption{
Kinematic distributions of some selected observables for the signal
with $M_N$ = 500 GeV (S, black with filled area), and for SM background processes of $t\bar{t}$ (red), WZ (blue), ZZ (cyan), and WWZ (green) after applying the pre-selection cuts at the FCC-hh.}
\label{fig:FCChh_distributions}
\end{figure}

\end{document}